\documentclass[12pt]{article}

\begin{document}

\hoffset=-0.75in
\voffset=-1.00in

\baselineskip=0.685cm
 
\title{Cosmological Perturbations in Conformal Gravity}

\author{Philip D. Mannheim \\Department of Physics,
University of Connecticut, Storrs, CT 06269, USA \\
e-mail: philip.mannheim@uconn.edu\\ }

\date{May 31, 2012}

\maketitle

We present the first steps needed for  an analysis of the perturbations that occur in the cosmology associated with the conformal gravity theory. We discuss the implications of conformal invariance for perturbative coordinate gauge choices, and show that in the conformal theory the trace of the metric fluctuation kinematically decouples from the first-order gravitational fluctuation equations. We determine the equations that describe first-order metric fluctuations around the illustrative conformal to flat  de Sitter background. Via a conformal transformation we show that  such fluctuations can be constructed from fluctuations around a flat background, even though the fluctuations themselves are associated with a perturbative geometry that is not itself conformal to flat. We extend the analysis to fluctuations around other cosmologically relevant backgrounds, such as the conformal to flat Robertson-Walker background, and find tensor fluctuations that grow far more rapidly than those that occur in the analogous standard case. We show that while the standard gravity tensor fluctuations around a de Sitter background are also fluctuation solutions  in the conformal theory, in the conformal case they  do not carry energy.

\section{Introduction}
\label{S1}

As a possible candidate alternative to standard Einstein gravity, conformal gravity is attractive in that it is a pure metric theory of gravity that possesses all of the general coordinate invariance and equivalence principle structure of standard gravity while augmenting it with an additional symmetry, local conformal invariance, in which  the action is left invariant under local conformal transformations on the metric of the form $g_{\mu\nu}(x)\rightarrow e^{2\alpha(x)}g_{\mu\nu}(x)$ with arbitrary local phase $\alpha(x)$. Under such a symmetry a gravitational action that is to be a polynomial function of the Riemann tensor is uniquely prescribed, and is given by (see e.g. \cite{Mannheim2006}) 
\begin{equation}
I_{\rm W}=-\alpha_g\int d^4x\, (-g)^{1/2}C_{\lambda\mu\nu\kappa}
C^{\lambda\mu\nu\kappa}
\equiv -2\alpha_g\int d^4x\, (-g)^{1/2}\left[R_{\mu\kappa}R^{\mu\kappa}-\frac{1}{3} (R^{\alpha}_{\phantom{\alpha}\alpha})^2\right],
\label{P1}
\end{equation}
where 
\begin{equation}
C_{\lambda\mu\nu\kappa}= R_{\lambda\mu\nu\kappa}
-\frac{1}{2}\left(g_{\lambda\nu}R_{\mu\kappa}-
g_{\lambda\kappa}R_{\mu\nu}-
g_{\mu\nu}R_{\lambda\kappa}+
g_{\mu\kappa}R_{\lambda\nu}\right)
+\frac{1}{6}R^{\alpha}_{\phantom{\alpha}\alpha}\left(
g_{\lambda\nu}g_{\mu\kappa}-
g_{\lambda\kappa}g_{\mu\nu}\right)
\label{P2}
\end{equation}
is the conformal Weyl tensor and the gravitational coupling constant $\alpha_g$ is 
dimensionless \cite{footnote0}. With the conformal symmetry forbidding the presence of any $\int d^4x\, (-g)^{1/2}\Lambda$ term in the action, the conformal theory has a control over the cosmological constant that the standard Einstein theory does not, and through this control one is able to address the cosmological constant problem 
\cite{Mannheim1990,Mannheim2008,Mannheim2009,Mannheim2010a}. Similarly, with the coupling constant $\alpha_g$ being dimensionless, unlike standard gravity, conformal gravity is power-counting renormalizable. And with its unitarity concerns having recently been 
addressed in \cite{Bender2008,Mannheim2009} (where a Hilbert space realization of the theory was constructed in which there are no negative norm states), and with its trace anomaly concerns having been addressed in \cite{Mannheim2011} (where it is shown that  the conformal trace anomaly in the matter sector can naturally be cancelled by the contribution of gravity itself  \cite{footnote00}), the theory is advanced \cite{Mannheim2011} as a potential quantum theory of gravity in four spacetime dimensions. Since conformal invariance requires that there be no intrinsic mass scales at the level of the Lagrangian, in the conformal theory all mass scales must come from the vacuum via spontaneous symmetry breaking. With such mass generation particles can then localize and bind into inhomogeneities such as the stars and galaxies that are of interest to astrophysics.

With the conformal theory being a renormalizable quantum theory at the microscopic level, then just as with electrodynamics, one is assured that its macroscopic classical predictions are reliable and not destroyed by quantum corrections. Consequently, application of the theory to macroscopic astrophysical phenomena allows one to test the theory. Early work in this direction was provided in \cite{Mannheim1997} where the theory was used to successfully fit the rotation curves of a set of 11 spiral galaxies, with the mass to light ratios ($M/L$) of the luminous optical disks of each of the galaxies being the only free parameters, and with no dark matter being required. More recently a systematic, broad-based study \cite{Mannheim2010b,Mannheim2010c,O'Brien2011} of rotation curves extended the fitting to a total of 138 galaxies (a varied sample consisting of high surface brightness, low surface brightness, and dwarf galaxies), and again acceptable fitting was obtained with the mass to light ratios of the galactic optical disks being the only free parameters, and again with no dark matter being required. The success of conformal gravity in fitting no less than 138 galactic rotation curves with just one free parameter per galaxy is very encouraging for the theory. And with its not needing to invoke dark matter, the very success of the fitting calls into question the validity of dark matter theory, where in addition to the optical disk mass to light ratios, one typically needs two free parameters for each galactic dark matter halo, and a thus additional 276 free parameters for the 138 galaxy sample. 

For cosmological applications, conformal gravity has been found capable of providing a natural explanation for the accelerating Universe supernovae data \cite{Mannheim2003,Mannheim2006}. Specifically, in the conformal theory the background Robertson-Walker (RW) cosmology is found to naturally be an accelerating one at all redshifts, so that unlike the situation that occurs in the standard cosmology, in the conformal case there is no need to fine-tune the theory so as to force the Universe to transition from deceleration to acceleration at a late redshift of order one. The conformal theory thus naturally addresses the dark energy problem that challenges the standard cosmological theory. 

In addition to galactic rotation curve data and cosmological background data, the conformal theory also needs to confront the wide variety of data (such as cluster dynamics, lensing by clusters, large scale structure, anisotropy of the cosmic microwave background) that are associated with departures from the homogeneity and isotropy of the cosmological background. In  this paper we take a first step in that direction. In Sec. II of this paper we discuss the general structure of the fluctuation equations in the gravitational sector of the conformal theory, and discuss the implications for these equations of conformal invariance and coordinate invariance. In Sec. III we determine the equations that describe first-order metric fluctuations around the illustrative conformal to flat de Sitter background, leading to our key (\ref{P49}) below. And in Sec. IV we utilize the underlying conformal structure of the conformal theory to relate the tensor mode (gravity wave) solutions of those equations to the tensor modes associated with fluctuations around a flat background, being able to do so even though neither set of tensor mode fluctuations is associated with a perturbative geometry that is itself conformal to flat. In Sec. V we extend our analysis to fluctuations around RW backgrounds with any possible spatial curvature $K$, and obtain 
closed-form expressions (\ref{P89}), (\ref{P115}) and (\ref{P117}) for the tensor mode fluctuations in the $K=0$, $K>0$ and  $K<0$ cases. In this section we also introduce the matter sector energy-momentum tensor $T^{\mu\nu}_{\rm M}$ needed to fix the expansion radius $a(t)$ of the RW geometry. Since the conformal gravity gravitational tensor (the quantity $W^{\mu\nu}$ given in (\ref{P3}) below) is traceless, in order for  $T^{\mu\nu}_{\rm M}$ to serve as its source, it too would need to be traceless. Now while a conformal invariant matter sector $T^{\mu\nu}_{\rm M}$ is automatically traceless in the symmetry limit in which particle masses are zero, in order for $T^{\mu\nu}_{\rm M}$ to continue to be the source of $W^{\mu\nu}$, it would need to remain traceless even after particles acquire masses and macroscopic inhomogeneities built out of them form. In Sec. V we show that the tracelessness of the matter sector energy-momentum tensor is in fact maintained when the matter fields acquire masses dynamically via a scalar field vacuum expectation value, with it being the energy and momentum of the selfsame scalar field that serves to keep the matter sector trace equal to zero. In Sec. VI we discuss some implications of our work, and in particular its connection to the rotation curve studies of  \cite{Mannheim2010b,Mannheim2010c,O'Brien2011}.  Also in Sec. VI  we show that while the standard gravity tensor fluctuations around a de Sitter background are also fluctuation solutions to conformal gravity, in the conformal case these particular fluctuations carry no energy. Finally, we note a possible connection between Einstein gravity and conformal gravity.

\section{Formalism and Coordinate and Conformal Invariance}
\label{S2}

\subsection{General Formalism}

With the Weyl action $I_{\rm W}$ given in  (\ref{P1}) being a fourth-order derivative function of the metric, functional variation with respect to the metric $g_{\mu\nu}(x)$ generates fourth-order derivative gravitational equations of motion of the form \cite{Mannheim2006} 
\begin{equation}
-\frac{2}{(-g)^{1/2}}\frac{\delta I_{\rm W}}{\delta g_{\mu\nu}}=4\alpha_g W^{\mu\nu}=4\alpha_g\left[
2C^{\mu\lambda\nu\kappa}_
{\phantom{\mu\lambda\nu\kappa};\lambda;\kappa}-
C^{\mu\lambda\nu\kappa}R_{\lambda\kappa}\right]=4\alpha_g\left[W^{\mu
\nu}_{(2)}-\frac{1}{3}W^{\mu\nu}_{(1)}\right]=T^{\mu\nu},
\label{P3}
\end{equation}
where the functions $W^{\mu \nu}_{(1)}$ and $W^{\mu \nu}_{(2)}$ are given by
\begin{eqnarray}
W^{\mu \nu}_{(1)}&=&
2g^{\mu\nu}(R^{\alpha}_{\phantom{\alpha}\alpha})          
^{;\beta}_{\phantom{;\beta};\beta}                                              
-2(R^{\alpha}_{\phantom{\alpha}\alpha})^{;\mu;\nu}                           
-2 R^{\alpha}_{\phantom{\alpha}\alpha}
R^{\mu\nu}                              
+\frac{1}{2}g^{\mu\nu}(R^{\alpha}_{\phantom{\alpha}\alpha})^2,
\nonumber\\
W^{\mu \nu}_{(2)}&=&
\frac{1}{2}g^{\mu\nu}(R^{\alpha}_{\phantom{\alpha}\alpha})   
^{;\beta}_{\phantom{;\beta};\beta}+
R^{\mu\nu;\beta}_{\phantom{\mu\nu;\beta};\beta}                     
 -R^{\mu\beta;\nu}_{\phantom{\mu\beta;\nu};\beta}                        
-R^{\nu \beta;\mu}_{\phantom{\nu \beta;\mu};\beta}                          
 - 2R^{\mu\beta}R^{\nu}_{\phantom{\nu}\beta}                                    
+\frac{1}{2}g^{\mu\nu}R_{\alpha\beta}R^{\alpha\beta}.
\label{P4}
\end{eqnarray}                                 
Since $W^{\mu\nu}$ is obtained from an action that is both general coordinate invariant and conformal invariant, it is kinematically covariantly conserved and traceless and obeys $W^{\mu\nu}_{\phantom{\mu\mu};\nu}=0$, $g_{\mu\nu}W^{\mu\nu}=0$.
   
Despite its somewhat formidable appearance, especially compared to the standard second-order derivative Einstein equations
\begin{equation}
-\frac{1}{8\pi G}\left(R^{\mu\nu} -\frac{1}{2}g^{\mu\nu}R^{\alpha}_{\phantom{\alpha}\alpha}\right)=T^{\mu\nu},
\label{P5}
\end{equation}
(\ref{P3}) immediately admits of two key vacuum solutions, namely solutions with vanishing Weyl tensor and solutions with vanishing Ricci tensor (and thus with vanishing $W^{\mu \nu}_{(1)}$ and $W^{\mu \nu}_{(2)}$). Solutions with vanishing Weyl tensor include the cosmologically relevant de Sitter and RW solutions, since the line elements of both geometries can be written in the conformal to flat (and thus vanishing Weyl tensor) form
\begin{equation}
ds^2=\Omega^2(t,x,y,z)[-dt^2+dx^2+dy^2+dz^2]
\label{P6}
\end{equation}
for appropriate choices of the conformal factor $\Omega(t,x,y,z)$. Solutions with vanishing Ricci tensor include all vacuum solutions to Einstein gravity such as the Schwarzschild solution exterior to a static, spherically symmetric source. 

In cosmological solutions the geometry is homogeneous and isotropic about every point of the spacetime,  to thereby cause the associated Weyl tensor to vanish, while the exterior Schwarzschild solution is only isotropic about a single point (the center of the spherical source) and has a non-vanishing exterior region Weyl tensor (see e.g. \cite{Mannheim1989}). The vanishing or non-vanishing of the Weyl tensor thus serves as a diagnostic for the homogeneity or otherwise of a geometry. Cosmological fluctuations around  conformaly flat geometries will generate non-conformally flat perturbative contributions to the metric and lead to the emergence of inhomogeneities such as the sources that can then serve to produce the Schwarzschild geometry. The objective of conformal gravity cosmological perturbation theory is thus to monitor the interplay between homogeneities and inhomogeneities, and in this paper we take a first step in that direction. We shall return to the issue of such an interplay in Sec. VI where we discuss the rotation curve studies of  \cite{Mannheim2010b,Mannheim2010c,O'Brien2011} from that perspective.

Because the Weyl action is locally conformal invariant, the function $W^{\mu\nu}(x)$ has the property that under 
\begin{equation}
g_{\mu\nu}(x)\rightarrow \Omega^2(x) g_{\mu\nu}(x)=\bar{g}_{\mu\nu}(x),\qquad
g^{\mu\nu}(x)\rightarrow \Omega^{-2}(x) g^{\mu\nu}(x)=\bar{g}^{\mu\nu}(x),
\label{P7}
\end{equation}
$W^{\mu\nu}(x)$ and $W_{\mu\nu}(x)$ transform as 
\begin{equation}
W^{\mu\nu}(x)\rightarrow \Omega^{-6}(x) W^{\mu\nu}(x)=\bar{W}^{\mu\nu}(x),\qquad
W_{\mu\nu}(x)\rightarrow \Omega^{-2}(x) W_{\mu\nu}(x)=\bar{W}_{\mu\nu}(x),
\label{P8}
\end{equation}
where the dependence of $\bar{W}_{\mu\nu}(x)$ on $\bar{g}_{\mu\nu}(x)$ is the same as that of 
$W_{\mu\nu}(x)$ on $g_{\mu\nu}(x)$. The great utility of (\ref{P8}) is that it holds regardless of whether or not the metric is conformal to flat. Moreover,  if we decompose each of $g_{\mu\nu}(x)$ and $\bar{g}_{\mu\nu}(x)$ into a background metric and a fluctuation according to
\begin{equation}
g_{\mu\nu}(x)=g^{(0)}_{\mu\nu}(x)+h_{\mu\nu}(x),\qquad
\bar{g}_{\mu\nu}(x)=\bar{g}^{(0)}_{\mu\nu}(x)+\bar{h}_{\mu\nu}(x),
\label{P9}
\end{equation}
then $W^{\mu\nu}(x)$ and $\bar{W}_{\mu\nu}(x)$ will decompose as 
\begin{equation}
W_{\mu\nu}(g_{\mu\nu})= W^{(0)}_{\mu\nu}(g^{(0)}_{\mu\nu})+\delta W_{\mu\nu}(h_{\mu\nu}),\qquad
\bar{W}_{\mu\nu}(\bar{g}_{\mu\nu})=\bar{W}^{(0)}_{\mu\nu}(\bar{g}^{(0)}_{\mu\nu})+\delta \bar{W}_{\mu\nu}(\bar{h}_{\mu\nu}),
\label{P10}
\end{equation}
where $W_{\mu\nu}(h_{\mu\nu})$ is evaluated in a background geometry with metric $g^{(0)}_{\mu\nu}(x)$, while $\bar{W}_{\mu\nu}(\bar{h}_{\mu\nu})$ is evaluated in a background geometry with metric $\bar{g}^{(0)}_{\mu\nu}(x)$. In addition, since the theory is conformal invariant, the matter sector $T_{\mu\nu}$ must transform as $\Omega^{-2}(x) T_{\mu\nu}(x)=\bar{T}_{\mu\nu}(x)$,  and decompose as
\begin{equation}
T_{\mu\nu}(g_{\mu\nu})= T^{(0)}_{\mu\nu}(g^{(0)}_{\mu\nu})+\delta T_{\mu\nu}(h_{\mu\nu}),\qquad
\bar{T}_{\mu\nu}(\bar{g}_{\mu\nu})=\bar{T}^{(0)}_{\mu\nu}(\bar{g}^{(0)}_{\mu\nu})+\delta \bar{T}_{\mu\nu}(\bar{h}_{\mu\nu}).
\label{P11}
\end{equation}
Thus if we know how to solve for fluctuations $h_{\mu\nu}(x)$ around a background $g^{(0)}_{\mu\nu}(x)$, i.e. if $g^{(0)}_{\mu\nu}(x)$ is such that we can actually find solutions to $\delta W_{\mu\nu}(h_{\mu\nu})=\delta T_{\mu\nu}(h_{\mu\nu})/4\alpha_g$, we can then obtain solutions to  $\delta \bar{W}_{\mu\nu}(\bar{h}_{\mu\nu})=\delta \bar{T}_{\mu\nu}(\bar{h}_{\mu\nu})/4\alpha_g$  for fluctuations $\bar{h}_{\mu\nu}(x)$ around a background metric $\bar{g}^{(0)}_{\mu\nu}(x)$ simply by setting 
\begin{equation}
\bar{h}_{\mu\nu}(x)=\Omega^2(x)h_{\mu\nu}(x).
\label{P12}
\end{equation}
Since the structure of the fluctuations around a flat background has already been obtained in \cite{Mannheim2009}, via. (\ref{P12}) we can construct the fluctuations around any background that is conformal to flat. Since all cosmologically relevant background geometries happen to be conformal to flat, this is extremely convenient, showing that despite its fourth-order derivative nature, there are simplifications in the conformal cosmological case that simply do not occur in standard second-order theory. To explicitly demonstrate the utility of (\ref{P12}), in Sec. III we shall calculate $\delta \bar{W}_{\mu\nu}(\bar{h}_{\mu\nu})$ directly in a background with a de Sitter $\bar{g}^{(0)}_{\mu\nu}(x)$ and then compare it with the $\delta W_{\mu\nu}(h_{\mu\nu})$ that is  obtained when $g^{(0)}_{\mu\nu}(x)$ is flat. 

For conformal to flat backgrounds there are yet other differences between the standard second-order theory and the conformal case. In the conformal theory case conformal to flat is a vacuum solution, to thus require that $T^{(0)}_{\mu\nu}(g^{(0)}_{\mu\nu})$ be zero. Since the standard theory does not take the energy-momentum tensor to be conformal invariant, it uses an energy-momentum tensor for which $T^{(0)}_{\mu\nu}(g^{(0)}_{\mu\nu})$ is non-zero, with the standard theory and the conformal theory approaches thus having to differ already in zeroth perturbative order. For the de Sitter background case for instance, the standard theory requires that the background energy-momentum tensor be given by $T^{(0)}_{\mu\nu}(g^{(0)}_{\mu\nu})=-\Lambda g^{(0)}_{\mu\nu}$ with cosmological constant $\Lambda$. Moreover, because such a background $T^{(0)}_{\mu\nu}(g^{(0)}_{\mu\nu})=-\Lambda g^{(0)}_{\mu\nu}$ is non-zero, in the presence of a perturbation the background geometry is modified and a perturbative contribution to the energy-momentum tensor of the form $\delta T_{\mu\nu}(h_{\mu\nu})=-\Lambda h_{\mu\nu}$ is induced. In the conformal case however, for conformally flat background geometries  there are no  induced perturbative terms of this type,  with the standard theory and the conformal theory thus having to differ not just in zeroth order but in first perturbative order as well. 

\subsection{Implications of Conformal Invariance}

For the purposes of actually evaluating the conformal gravity fluctuation term  $\delta W_{\mu\nu}(h_{\mu\nu})$, there are some convenient simplifications that arise due to the conformal invariance and coordinate invariance of the theory. As regards first simplifications due to conformal invariance, consider a situation in which there is just a background metric $g^{(0)}_{\mu\nu}(x)$ and no explicit fluctuation, with the background metric being such that $W^{(0)}_{\mu\nu}(g^{(0)}_{\mu\nu})=0$. Now make  a particular choice of conformal transformation of the form $\Omega^2(x)=(1+A/4)$, where for the moment $A$ is just an arbitrary small quantity. Under this transformation the background metric transforms as 
\begin{equation}
g^{(0)}_{\mu\nu}(x) \rightarrow g^{(0)}_{\mu\nu}(x)+\frac{A}{4}g^{(0)}_{\mu\nu}(x)
\label{P13}
\end{equation}
while $W^{(0)}_{\mu\nu}(g^{(0)}_{\mu\nu})$ transforms as 
\begin{equation}
W^{(0)}_{\mu\nu}(g^{(0)}_{\mu\nu})\rightarrow \left[ 1- \frac{A}{4}\right]W^{(0)}_{\mu\nu}(g^{(0)}_{\mu\nu})
\label{P14}
\end{equation}
(as evaluated to lowest order in $A$),  and thus remains zero. From the perspective of perturbation theory we can interpret (\ref{P13}) as a perturbation on the background metric of the form $h_{\mu\nu}=Ag^{(0)}_{\mu\nu}/4$, and can interpret (\ref{P14}) as a perturbation on the gravitational equations of motion of the form $W_{\mu\nu}(g_{\mu\nu})=W^{(0)}_{\mu\nu}(g^{(0)}_{\mu\nu})+\delta{W}_{\mu\nu}(h_{\mu\nu})$. Since $W^{(0)}_{\mu\nu}(g^{(0)}_{\mu\nu})$ has been set to zero, it follows from (\ref{P14}) that when evaluated with an $h_{\mu\nu}$ given by $h_{\mu\nu}=Ag^{(0)}_{\mu\nu}/4$, the quantity $\delta W_{\mu\nu}(Ag^{(0)}_{\mu\nu}/4)$ must vanish identically. However, since the trace of any $h_{\mu\nu}$ as determined with respect to the background metric is given by $h=g_{(0)}^{\mu\nu}h_{\mu\nu}$, it follows that for the particular  $h_{\mu\nu}$ given by $h_{\mu\nu}=Ag^{(0)}_{\mu\nu}/4$, its trace is given by $h=A$. Thus when there is an explicit  fluctuation $h_{\mu\nu}$, it follows that since there can be no cross-terms between $h_{\mu\nu}$ and its trace in first order, the first-order perturbative $\delta W_{\mu\nu}(hg^{(0)}_{\mu\nu}/4)$, and thus the first-order $\delta\bar{W}_{\mu\nu}(\bar{h}\bar{g}^{(0)}_{\mu\nu}/4)$, must vanish identically. Consequently, the trace of the fluctuation can make no contribution to the first-order gravitational fluctuation equations. Thus even though it need not vanish, the trace does not make any contribution to $\delta W_{\mu\nu}(h_{\mu\nu})$.

To take advantage of this decoupling we introduce a quantity $K_{\mu\nu}(x)$ defined as 
\begin{equation}
K_{\mu\nu}(x)=h_{\mu\nu}(x)-\frac{1}{4}g^{(0)}_{\mu\nu}(x)g_{(0)}^{\alpha\beta}h_{\alpha\beta},
\label{P15}
\end{equation}
with $K_{\mu\nu}$ being traceless with respect to the background metric $g_{(0)}^{\mu\nu}$. If we now evaluate $\delta W_{\mu\nu}(h_{\mu\nu})=\delta W_{\mu\nu}(K_{\mu\nu}+hg^{(0)}_{\mu\nu}/4)$ for some general fluctuation $h_{\mu\nu}$, we will find that the dependence on $h$ will drop out identically and $\delta W_{\mu\nu}(h_{\mu\nu})$ will be given as $\delta W_{\mu\nu}(h_{\mu\nu})=\delta W_{\mu\nu}(K_{\mu\nu})$. Rather than be a function of the 10-component $h_{\mu\nu}$, $\delta W_{\mu\nu}$ must instead be a function of the 9-component $K_{\mu\nu}$ alone. Note that we are not asserting here that $h_{\mu\nu}$ has been made traceless by a conformal transformation (in fact it could not be since $g_{(0)}^{\mu\nu}h_{\mu\nu}$ is conformal invariant). Rather, we are asserting that the first-order fluctuation in $\delta W_{\mu\nu}$ can only depend on the traceless combination $K_{\mu\nu}=h_{\mu\nu}-g^{(0)}_{\mu\nu}h/4$ rather than on $h_{\mu\nu}$ itself, an extremely convenient simplification. In \cite{Mannheim2006,Mannheim2009} we had already found this to explicitly be the case for perturbations around flat spacetime, and in Sec. III we shall explicitly demonstrate it by brute force for fluctuations around a de Sitter background.  

Because of the decoupling of the trace,  we can replace (\ref{P10}) by 
\begin{equation}
W_{\mu\nu}(g_{\mu\nu})= W^{(0)}_{\mu\nu}(g^{(0)}_{\mu\nu})+\delta W_{\mu\nu}(K_{\mu\nu}),\qquad
\bar{W}_{\mu\nu}(\bar{g}_{\mu\nu})=\bar{W}^{(0)}_{\mu\nu}(\bar{g}^{(0)}_{\mu\nu})+\delta\bar{W}_{\mu\nu}(\bar{K}_{\mu\nu}),
\label{P16}
\end{equation}
where
\begin{equation}
\bar{g}^{(0)}_{\mu\nu}(x)=\Omega^2(x)g^{(0)}_{\mu\nu}(x),
\label{P17}
\end{equation}
\begin{equation}
\bar{K}_{\mu\nu}(x)=\Omega^2(x)K_{\mu\nu}(x).
\label{P18}
\end{equation}
In the following then, to construct the fluctuations in a $\bar{g}^{(0)}_{\mu\nu}$ background from the fluctuations in a $g^{(0)}_{\mu\nu}$ background, we shall need to utilize (\ref{P18}) rather than (\ref{P12}). 

The treatment of the trace of the fluctuation in the conformal case is totally differently from the way the trace is treated in the second-order theory. In the second-order theory one starts with a 10-component $h_{\mu\nu}$ and via a sequence of four coordinate transformations, and without reference to the fluctuation equations,  one reduces to six independent fluctuation components, with the trace of $h_{\mu\nu}$ not being brought to zero (c.f. the typical harmonic gauge choice $\partial_{\nu}h^{\mu\nu}=\partial^{\mu}h/2$). It is only after one subsequently imposes a set of four  additional residual gauge symmetries that solutions to the fluctuation equations possess that one is then able to reduce to just two independent components, with the trace $h$ being zero in solutions to the fluctuation equations, and with the solutions then being both transverse and traceless. In contrast, in the conformal theory one can reduce the theory to a dependence on the traceless $K_{\mu\nu}$ without needing to make any reference to the fluctuation equations at all. Since one also has the freedom to make four general coordinate transformations, on using them one can reduce  the 9-component $K_{\mu\nu}$ to five independent components, again without needing to make any reference to the fluctuation equations. Any further reduction in the number of independent components of $K_{\mu\nu}$ can only be achieved through use of residual gauge invariances or the structure of the  fluctuation equations themselves.

\subsection{Implications of Coordinate Invariance}

As regards the four general coordinate transformations, it would be very convenient if we could arrange for $K_{\mu\nu}$ to be transverse, since before even beginning to consider the fluctuation equations, we would then have reduced to five independent components, with all of them being transverse and traceless.  Thus not only is the trace treated differently in the conformal theory, transverseness is treated differently too. However, in trying to impose transverseness in the form $K^{\mu\nu}_{\phantom{\mu\nu};\nu}=0$, we run into an initial difficulty, namely that the $K^{\mu\nu}_{\phantom{\mu\nu};\nu}=0$ condition is not preserved under a conformal transformation. 

To be specific, we note that in a $g^{(0)}_{\mu\nu}$ background, $K^{\mu\nu}_{\phantom{\mu\nu};\nu}$ takes the form
\begin{equation}
K^{\mu\nu}_{\phantom{\mu\nu};\nu}=
\partial_{\nu}K^{\mu\nu}
+K^{\nu\sigma}g_{(0)}^{\mu\rho}\partial_{\nu}g^{(0)}_{\rho\sigma}
-\frac{1}{2}K^{\nu\sigma}g_{(0)}^{\mu\rho}\partial_{\rho}g^{(0)}_{\nu\sigma}
+\frac{1}{2}K^{\mu\sigma}g_{(0)}^{\nu\rho}\partial_{\sigma}g^{(0)}_{\rho\nu}.
\label{P19}
\end{equation}
On recalling that $K^{\nu\sigma}g^{(0)}_{\nu\sigma}=0$, we find that under a conformal transformation $K^{\mu\nu}_{\phantom{\mu\nu};\nu}$ transforms as
\begin{equation}
K^{\mu\nu}_{\phantom{\mu\nu};\nu}\rightarrow \Omega^{-2}K^{\mu\nu}_{\phantom{\mu\nu};\nu}
+4\Omega^{-3}K^{\mu\sigma}\partial_{\sigma}\Omega,
\label{P20}
\end{equation}
to thus not be conformal invariant. To identify a coordinate gauge condition that is conformal invariant, we note that under a conformal transformation the quantity $K^{\mu\nu}g_{(0)}^{\alpha\beta}\partial_{\nu}g^{(0)}_{\alpha\beta}$ transforms as
\begin{equation}
K^{\mu\nu}g_{(0)}^{\alpha\beta}\partial_{\nu}g^{(0)}_{\alpha\beta}
\rightarrow \Omega^{-2}K^{\mu\nu}g_{(0)}^{\alpha\beta}\partial_{\nu}g^{(0)}_{\alpha\beta}
+8\Omega^{-3}K^{\mu\nu}\partial_{\nu}\Omega.
\label{P21}
\end{equation}
Consequently, we obtain
\begin{equation}
K^{\mu\nu}_{\phantom{\mu\nu};\nu}-\frac{1}{2} K^{\mu\nu}g_{(0)}^{\alpha\beta}\partial_{\nu}g^{(0)}_{\alpha\beta}\rightarrow \Omega^{-2}\left[K^{\mu\nu}_{\phantom{\mu\nu};\nu}
-\frac{1}{2} K^{\mu\nu}g_{(0)}^{\alpha\beta}\partial_{\nu}g^{(0)}_{\alpha\beta}\right]
=\overline{K^{\mu\nu}_{\phantom{\mu\nu};\nu}}-\frac{1}{2} \bar{K}^{\mu\nu}\bar{g}_{(0)}^{\alpha\beta}\partial_{\nu}\bar{g}^{(0)}_{\alpha\beta},
\label{P22}
\end{equation}
where $\overline{K^{\mu\nu}_{\phantom{\mu\nu};\nu}}$ is evaluated in a geometry with metric $\bar{g}^{(0)}_{\mu\nu}$ according to 
\begin{equation}
\overline{K^{\mu\nu}_{\phantom{\mu\nu};\nu}}=
\partial_{\nu}\bar{K}^{\mu\nu}
+\bar{K}^{\nu\sigma}\bar{g}_{(0)}^{\mu\rho}\partial_{\nu}\bar{g}^{(0)}_{\rho\sigma}
-\frac{1}{2}\bar{K}^{\nu\sigma}\bar{g}_{(0)}^{\mu\rho}\partial_{\rho}\bar{g}^{(0)}_{\nu\sigma}
+\frac{1}{2}\bar{K}^{\mu\sigma}\bar{g}_{(0)}^{\nu\rho}\partial_{\sigma}\bar{g}^{(0)}_{\rho\nu}.
\label{P23}
\end{equation}

If we wish to use the conformal transformation given in (\ref{P18}) to construct $\bar{K}_{\mu\nu}$ from $K_{\mu\nu}$, from (\ref{P22}) we see that if we were to work in the coordinate gauge $K^{\mu\nu}_{\phantom{\mu\nu};\nu}- K^{\mu\nu}g_{(0)}^{\alpha\beta}\partial_{\nu}g^{(0)}_{\alpha\beta}/2=0$, we would then conformally transform into the coordinate gauge where $\overline{K^{\mu\nu}_{\phantom{\mu\nu};\nu}}-\bar{K}^{\mu\nu}\bar{g}_{(0)}^{\alpha\beta}\partial_{\nu}\bar{g}^{(0)}_{\alpha\beta}/2=0$. Since it would be preferable to solve an equation such as $\delta W_{\mu\nu}(K_{\mu\nu})=0$ in a straightforward gauge such as $K^{\mu\nu}_{\phantom{\mu\nu};\nu}=0$, should we choose to do this, then prior to making the conformal transformation we  would first have to make a perturbative gauge transformation to the coordinate gauge with $K^{\mu\nu}_{\phantom{\mu\nu};\nu}- K^{\mu\nu}g_{(0)}^{\alpha\beta}\partial_{\nu}g^{(0)}_{\alpha\beta}/2=0$. Recalling that  $h_{\mu\nu}\rightarrow h_{\mu\nu}+\epsilon_{\mu;\nu}+\epsilon_{\nu;\mu}$ under a perturbative coordinate gauge transformation, we see that under the same transformation $K_{\mu\nu}$ would transform as  $K_{\mu\nu}\rightarrow K_{\mu\nu}+\epsilon_{\mu;\nu}+\epsilon_{\nu;\mu}-g^{(0)}_{\mu\nu}\epsilon^{\alpha}_{\phantom{\alpha};\alpha}/2$. We would thus need to pick the $\epsilon_{\mu}$ so as to effect a transformation from the coordinate gauge with $K^{\mu\nu}_{\phantom{\mu\nu};\nu}=0$ to that with $K^{\mu\nu}_{\phantom{\mu\nu};\nu}- K^{\mu\nu}g_{(0)}^{\alpha\beta}\partial_{\nu}g^{(0)}_{\alpha\beta}/2=0$. Then we would have to make a conformal transformation to take us into the gauge with $\overline{K^{\mu\nu}_{\phantom{\mu\nu};\nu}}-\bar{K}^{\mu\nu}\bar{g}_{(0)}^{\alpha\beta}\partial_{\nu}\bar{g}^{(0)}_{\alpha\beta}/2=0$, and finally we would have to make yet another coordinate gauge transformation to get to the gauge with $\overline{K^{\mu\nu}_{\phantom{\mu\nu};\nu}}=0$.

We could reduce the number of steps involved by rewriting (\ref{P22}) as
\begin{equation}
 \Omega^{-2}K^{\mu\nu}_{\phantom{\mu\nu};\nu}
=\overline{K^{\mu\nu}_{\phantom{\mu\nu};\nu}}-\frac{1}{2} \bar{K}^{\mu\nu}\bar{g}_{(0)}^{\alpha\beta}\partial_{\nu}\bar{g}^{(0)}_{\alpha\beta}
+\frac{1}{2} \Omega^{-2}K^{\mu\nu}g_{(0)}^{\alpha\beta}\partial_{\nu}g^{(0)}_{\alpha\beta}
=\overline{K^{\mu\nu}_{\phantom{\mu\nu};\nu}}-4\Omega^{-1}\bar{K}^{\mu\nu}\partial_{\nu}\Omega.
\label{P24}
\end{equation}
Then we could start in the coordinate gauge with $K^{\mu\nu}_{\phantom{\mu\nu};\nu}=0$, conformally transform into the coordinate gauge with $\overline{K^{\mu\nu}_{\phantom{\mu\nu};\nu}}-4\Omega^{-1}\bar{K}^{\mu\nu}\partial_{\nu}\Omega=0$, and finally only need to make one coordinate gauge transformation to get to the coordinate gauge in which $\overline{K^{\mu\nu}_{\phantom{\mu\nu};\nu}}=0$.

While the above sequence of coordinate and conformal transformations would be needed for the general background metric, great simplification would be obtained if we could eliminate the $\Omega^{-1}\bar{K}^{\mu\nu}\partial_{\nu}\Omega$ term from (\ref{P24}). And this can actually be achieved if the $\bar{g}^{\mu\nu}_{(0)}$ background metric is conformal to a flat $g^{(0)}_{\mu\nu}$ but with a conformal factor that only depends on the time coordinate. Specifically, in that case for those particular modes that obey the synchronous mode condition  $\bar{K}_{0\mu}=0$, we find that the $\Omega^{-1}\bar{K}^{\mu\nu}\partial_{\nu}\Omega$ term would then vanish identically, with (\ref{P24}) simplifying to
\begin{equation}
 \Omega^{-2}K^{\mu\nu}_{\phantom{\mu\nu};\nu}
=\overline{K^{\mu\nu}_{\phantom{\mu\nu};\nu}},
\label{P25}
\end{equation}
and with the transverse derivatives then transforming into each other under a conformal transformation. (If the conformal factor depended on just one spatial coordinate, $z$ say, we could instead work in the axial gauge where $K_{3\mu}=0$.)  Now conformal to flat metrics of the form given in (\ref{P6}) with a conformal factor that only depends on the time coordinate are of great practical interest since they include both de Sitter and spatially flat RW cosmologies. (The cases with a conformal factor that instead depends on $z$ include anti de Sitter space). As we shall now show, in the de Sitter case the structure of the fluctuation equations is such that for the tensor mode (gravity wave) solutions that we study below we can simultaneously maintain both $\overline{K^{\mu\nu}_{\phantom{\mu\nu};\nu}}=0$ and $\bar{K}_{0\mu}=0$, the former because of the gauge freedom of the general $\delta \bar{W}_{\mu\nu}(\bar{K}_{\mu\nu})$, and the latter because of the structure of solutions to $\delta \bar{W}_{\mu\nu}(\bar{K}_{\mu\nu})=0$.

\section{Fluctuations Around a de Sitter Background}
\label{S3}

\subsection{Preliminaries}

In a background de Sitter geometry with metric $g_{\mu\nu}(x)$ the Riemann tensor, the Ricci tensor and the Ricci scalar are given by
\begin{equation}
R_{\kappa\nu\rho\sigma}=H^2(g_{\rho\nu}g_{\kappa\sigma}-g_{\sigma\nu}g_{\kappa\rho}),\qquad R_{\nu\sigma}=-3H^2g_{\nu\sigma},\qquad R^{\alpha}_{\phantom{\alpha}\alpha}=-12H^2,
\label{P26}
\end{equation}
where $H$ is a constant. Inspection of (\ref{P4}) shows that to evaluate the fluctuation term $\delta W_{\mu\nu}(h_{\mu\nu})$ in a fluctuation $h_{\mu\nu}(x)$ around a $g_{\mu\nu}(x)$ background, the primary fluctuation term we will need to determine is $\delta [\nabla_{\sigma}\nabla_{\lambda}R_{\mu\nu}]$. On recalling that 
\begin{equation}
\nabla_{\sigma}\nabla_{\lambda}R_{\mu\nu}=\partial_{\sigma}[\nabla_{\lambda}R_{\mu\nu}]
-\Gamma^{\rho}_{\sigma\mu}\nabla_{\lambda}R_{\rho\nu}
-\Gamma^{\rho}_{\sigma\nu}\nabla_{\lambda}R_{\mu\rho}
-\Gamma^{\rho}_{\sigma\lambda}\nabla_{\rho}R_{\mu\nu},
\label{P27}
\end{equation}
we see that since the $\nabla_{\lambda}R_{\rho\nu}$, $ \nabla_{\lambda}R_{\mu\rho}$ and $\nabla_{\rho}R_{\mu\nu}$ terms all vanish in a geometry in which the Ricci tensor is proportional to the metric tensor, fluctuation contributions to any of the Christoffel symbol terms in (\ref{P27}) make no contribution to $\delta [\nabla_{\sigma}\nabla_{\lambda}R_{\mu\nu}]$. Thus we can set $\delta [\nabla_{\sigma}\nabla_{\lambda}R_{\mu\nu}]=
\nabla_{\sigma}[\delta [\nabla_{\lambda}R_{\mu\nu}]]$. Recalling next that 
\begin{equation}
\nabla_{\lambda}R_{\mu\nu}=\partial_{\lambda}R_{\mu\nu}
-\Gamma^{\rho}_{\lambda\mu}R_{\rho\nu}
-\Gamma^{\rho}_{\lambda\nu}R_{\rho\mu},
\label{P28}
\end{equation}
we obtain
\begin{equation}
\delta[\nabla_{\lambda}R_{\mu\nu}]=\nabla_{\lambda}[\delta R_{\mu\nu}]
-\delta[\Gamma^{\rho}_{\lambda\mu}]R_{\rho\nu}
-\delta[\Gamma^{\rho}_{\lambda\nu}]R_{\rho\mu}.
\label{P29}
\end{equation}
On recalling that
\begin{equation}
\delta[\Gamma^{\rho}_{\lambda\mu}]=\frac{1}{2}g^{\rho\sigma}[\nabla_{\lambda}h_{\sigma\mu}
+\nabla_{\mu}h_{\sigma\lambda}-\nabla_{\sigma}h_{\lambda\mu}],
\label{P30}
\end{equation}
and using (\ref{P26}), we obtain
\begin{equation}
\delta[\nabla_{\lambda}R_{\mu\nu}]=\nabla_{\lambda}[\delta R_{\mu\nu}]
+3H^2\nabla_{\lambda}h_{\mu\nu}. 
\label{P31}
\end{equation}
Consequently the general $\delta [\nabla_{\sigma}\nabla_{\lambda}R_{\mu\nu}]$ fluctuation term is given by 
\begin{equation}
\delta [\nabla_{\sigma}\nabla_{\lambda}R_{\mu\nu}]=\nabla_{\sigma}\nabla_{\lambda}[\delta R_{\mu\nu}+3H^2h_{\mu\nu}].
\label{P32}
\end{equation}

The fluctuation quantity $\delta R_{\mu\kappa}$ that appears in (\ref{P32}) is of the form 
\begin{equation}
\delta R_{\mu\kappa}=\frac{1}{2}g^{\lambda \rho}[\nabla_{\kappa}\nabla_{\mu}h_{\lambda\rho}
-\nabla_{\lambda}\nabla_{\kappa}h_{\rho\mu}
-\nabla_{\lambda}\nabla_{\mu}h_{\rho\kappa}
+\nabla_{\lambda}\nabla_{\rho}h_{\mu\kappa}].
\label{P33}
\end{equation}
On recalling that for any rank 2 tensor
\begin{equation}
\nabla_{\kappa}\nabla_{\nu}A_{\lambda\mu}
-\nabla_{\nu}\nabla_{\kappa}A_{\lambda\mu}
=A^{\sigma}_{\phantom{\sigma}\mu}R_{\lambda\sigma\nu\kappa}
-A_{\lambda}^{\phantom{\lambda}\sigma}R_{\sigma\mu\nu\kappa},
\label{P34}
\end{equation}
we can write $\delta R_{\mu\kappa}$ as 
\begin{eqnarray}
\delta R_{\mu\kappa}&=&\frac{1}{2}g^{\lambda \rho}[\nabla_{\kappa}\nabla_{\mu}h_{\lambda\rho}
-\nabla_{\kappa}\nabla_{\lambda}h_{\rho\mu}
-\nabla_{\mu}\nabla_{\lambda}h_{\rho\kappa}
+\nabla_{\lambda}\nabla_{\rho}h_{\mu\kappa}
\nonumber\\
&-&h^{\sigma}_{\phantom{\sigma}\mu}R_{\rho\sigma\kappa\lambda}
+h^{\sigma}_{\phantom{\sigma}\rho}R_{\sigma\mu\kappa\lambda}
-h^{\sigma}_{\phantom{\sigma}\kappa}R_{\rho\sigma\mu\lambda}
+h^{\sigma}_{\phantom{\sigma}\rho}R_{\sigma\kappa\mu\lambda}].
\label{P35}
\end{eqnarray}
Thus given (\ref{P26}), we find that  $\delta R_{\mu\kappa}$ and $g^{\mu\kappa}\delta R_{\mu\kappa}$ take the very convenient forms 
\begin{eqnarray}
\delta R_{\mu\kappa}&=&\frac{1}{2}[\nabla_{\kappa}\nabla_{\mu}h
-\nabla_{\kappa}\nabla_{\lambda}h^{\lambda}_{\phantom{\lambda}\mu}
-\nabla_{\mu}\nabla_{\lambda}h^{\lambda}_{\phantom{\lambda}\kappa}
+\nabla_{\lambda}\nabla^{\lambda}h_{\mu\kappa}
-8H^2h_{\mu\kappa}+2H^2hg_{\mu\kappa}],
\nonumber\\
g^{\mu\kappa}\delta R_{\mu\kappa}&=&\nabla_{\lambda}\nabla^{\lambda}h-\nabla_{\kappa}\nabla_{\lambda}h^{\kappa\lambda}.
\label{P36}
\end{eqnarray}
Armed with (\ref{P32}) and (\ref{P36}) and recalling that for a covariant fluctuation $\delta g_{\mu\nu}=h_{\mu\nu}$ the contravariant fluctuation is given by $\delta g^{\mu\nu}=-h^{\mu\nu}$, we can now proceed to evaluate $\delta W_{\mu\nu}(h_{\mu\nu})$.

\subsection{Evaluation of $\delta W^{(1)}_{\mu\nu}$}

Evaluating first $\delta W^{(1)}_{\mu\nu}(h_{\mu\nu})$, from (\ref{P4}) we directly obtain 
\begin{eqnarray}
\delta W^{(1)}_{\mu\nu}&=&2g^{\sigma\tau}[g_{\mu\nu}\nabla_{\beta}\nabla^{\beta}
-\nabla_{\nu}\nabla_{\mu}][\delta R_{\sigma\tau}+3H^2h_{\sigma\tau}]
+2h^{\sigma\tau}R_{\sigma\tau}R_{\mu\nu}-2g^{\sigma\tau}\delta R_{\sigma\tau}R_{\mu\nu}
\nonumber\\
&-&2R^{\alpha}_{\phantom{\alpha}\alpha}\delta R_{\mu\nu}
+\frac{1}{2}h^{\mu\nu}(R^{\alpha}_{\phantom{\alpha}\alpha})^2
+g_{\mu\nu}R^{\alpha}_{\phantom{\alpha}\alpha}[
g^{\sigma\tau}\delta R_{\sigma\tau}-h^{\sigma\tau}R_{\sigma\tau}],
\label{P37}
\end{eqnarray}
and thus
\begin{eqnarray}
\delta W^{(1)}_{\mu\nu}&=&[2g_{\mu\nu}\nabla_{\alpha}\nabla^{\alpha} -2\nabla_{\mu}\nabla_{\nu}
-6H^2g_{\mu\nu}]
[\nabla_{\beta}\nabla^{\beta}h-\nabla_{\lambda}\nabla_{\kappa}h^{\lambda\kappa}+3H^2h]
+72H^4h_{\mu\nu}
\nonumber\\
&+&12H^2[\nabla_{\mu}\nabla_{\nu}h
-\nabla_{\mu}\nabla_{\lambda}h^{\lambda}_{\phantom{\lambda}\nu}
-\nabla_{\nu}\nabla_{\lambda}h^{\lambda}_{\phantom{\lambda}\mu}
+\nabla_{\lambda}\nabla^{\lambda}h_{\mu\nu}
-8H^2h_{\mu\nu}+2H^2g_{\mu\nu}h].
\label{P38}
\end{eqnarray}
On making the substitution $h_{\mu\nu}=K_{\mu\nu}+g_{\mu\nu}h/4$ as per (\ref{P15}), we find that  (\ref{P38}) takes the form  
\begin{eqnarray}
\delta W^{(1)}_{\mu\nu}&=&-[2g_{\mu\nu}\nabla_{\alpha}\nabla^{\alpha} -2\nabla_{\mu}\nabla_{\nu}
-6H^2g_{\mu\nu}]\nabla_{\lambda}\nabla_{\kappa}K^{\lambda\kappa}
\nonumber\\
&+&12H^2[\nabla_{\lambda}\nabla^{\lambda}K_{\mu\nu}
-2H^2K_{\mu\nu}
-\nabla_{\mu}\nabla_{\lambda}K^{\lambda}_{\phantom{\lambda}\nu}
-\nabla_{\nu}\nabla_{\lambda}K^{\lambda}_{\phantom{\lambda}\mu}]
\nonumber\\
&+&\frac{3}{2}g_{\mu\nu}\nabla_{\alpha}\nabla^{\alpha}\nabla_{\beta}\nabla^{\beta}h
-\frac{3}{2}\nabla_{\mu}\nabla_{\nu}\nabla_{\alpha}\nabla^{\alpha}h
+\frac{9}{2}H^2g_{\mu\nu}\nabla_{\alpha}\nabla^{\alpha}h,
\label{P39}
\end{eqnarray}
with the dependence of $\delta W^{(1)}_{\mu\nu}$ on $K_{\mu\nu}$ and $h$ in a de Sitter background being nicely organized.

\subsection{Evaluation of $\delta W^{(2)}_{\mu\nu}$}

Evaluating now  $\delta W^{(2)}_{\mu\nu}(h_{\mu\nu})$, from (\ref{P4}) we directly obtain 
\begin{eqnarray}
\delta W^{(2)}_{\mu\nu}&=&\frac{1}{2}g^{\sigma\tau}g_{\mu\nu}\nabla_{\beta}\nabla^{\beta}
[\delta R_{\sigma\tau}+3H^2h_{\sigma\tau}]
+\nabla_{\beta}\nabla^{\beta}[\delta R_{\mu\nu}+3H^2h_{\mu\nu}]
\nonumber\\
&-&g^{\beta\sigma}\nabla_{\sigma}\nabla_{\nu}[
\delta R_{\mu\beta}+3H^2h_{\mu\beta}]
-g^{\beta\sigma}\nabla_{\sigma}\nabla_{\mu}[
\delta R_{\nu\beta}+3H^2h_{\nu\beta}]
\nonumber\\
&+&2h^{\alpha\beta}R_{\mu\alpha}R_{\nu\beta}
-2g^{\alpha\beta}\delta R_{\mu\alpha}R_{\nu\beta}
-2g^{\alpha\beta}R_{\mu\alpha}\delta R_{\nu\beta}
+\frac{1}{2}h_{\mu\nu}R_{\alpha\beta}R^{\alpha\beta}
\nonumber\\
&+&\frac{1}{2}g_{\mu\nu}\delta R_{\alpha\beta}R^{\alpha\beta}
+\frac{1}{2}g_{\mu\nu}R_{\alpha\beta}[-(h^{\alpha\sigma}g^{\beta\tau}+g^{\alpha\sigma}h^{\beta\tau})R_{\sigma\tau}+g^{\alpha\sigma}g^{\beta\tau}\delta R_{\sigma\tau}].
\label{P40}
\end{eqnarray}
Since from (\ref{P34}) we obtain
\begin{equation}
g^{\beta\sigma}[\nabla_{\sigma}\nabla_{\nu}A_{\beta \mu}
-\nabla_{\nu}\nabla_{\sigma}A_{\beta\mu}]
=4H^2A_{\nu\mu}-H^2g_{\mu\nu}g^{\alpha\beta}A_{\alpha\beta}
\label{P41}
\end{equation}
for any rank 2 tensor in a de Sitter background, we can rewrite (\ref{P40}) as
\begin{eqnarray}
\delta W^{(2)}_{\mu\nu}&=&[\nabla_{\beta}\nabla^{\beta}+4H^2]\delta R_{\mu\nu}
-\nabla_{\nu}\nabla^{\beta}\delta R_{\mu\beta}-\nabla_{\mu}\nabla^{\beta}\delta R_{\nu\beta}
\nonumber\\
&+&3H^2\nabla_{\beta}\nabla^{\beta}h_{\mu\nu}
-3H^2\nabla_{\nu}\nabla^{\beta}h_{\mu\beta}
-3H^2\nabla_{\mu}\nabla^{\beta}h_{\nu\beta}
+12H^4h_{\mu\nu}-3H^4g_{\mu\nu}h
\nonumber\\
&+&\frac{1}{2}g_{\mu\nu}[\nabla_{\beta}\nabla^{\beta}-2H^2][\nabla_{\lambda}\nabla^{\lambda}h-\nabla_{\kappa}\nabla_{\lambda}h^{\kappa\lambda}]
+\frac{3}{2}g_{\mu\nu}H^2\nabla_{\beta}\nabla^{\beta}h.
\label{P42}
\end{eqnarray}

It is possible to simplify (\ref{P42}) using the perturbative Bianchi identity for the Einstein tensor  viz. $\delta[\nabla^{\nu}G_{\mu\nu}]=-h^{\sigma\nu}\nabla_{\sigma}G_{\mu\nu}+g^{\sigma\nu}\delta[\nabla_{\sigma}G_{\mu\nu}]=0$, where $G_{\mu\nu}=R_{\mu\nu}-g_{\mu\nu} g^{\alpha\beta}R_{\alpha\beta}/2$ and
\begin{eqnarray}
\delta[\nabla_{\sigma}G_{\mu\nu}]&=&\nabla_{\sigma}\delta[G_{\mu\nu}]
-\delta \Gamma^{\rho}_{\sigma\mu}G_{\rho\nu}
-\delta \Gamma^{\rho}_{\sigma\nu}G_{\rho\mu}
\nonumber\\
&=&\nabla_{\sigma}\delta[R_{\mu\nu}]
-\frac{1}{2}\nabla_{\sigma}[h_{\mu\nu}g^{\alpha\beta}R_{\alpha\beta}
-g_{\mu\nu}h^{\alpha\beta}R_{\alpha\beta}+g_{\mu\nu}g^{\alpha\beta}\delta R_{\alpha\beta}]
\nonumber\\
&-&\frac{1}{2}G_{\rho\nu}g^{\rho\tau}[\nabla_{\mu}h_{\tau\sigma}+\nabla_{\sigma}h_{\tau\mu} -\nabla_{\tau}h_{\mu\sigma}]
-\frac{1}{2}G_{\rho\mu}g^{\rho\tau}[\nabla_{\nu}h_{\tau\sigma}+\nabla_{\sigma}h_{\tau\nu} -\nabla_{\tau}h_{\nu\sigma}].
\nonumber\\
\label{P43}
\end{eqnarray}
Thus in a de Sitter background where  the $G_{\mu\nu}=3H^2g_{\mu\nu}$ form for $G_{\mu\nu}$ entails that $\nabla_{\sigma}G_{\mu\nu}=0$ and thus that $g^{\sigma\nu}\delta[\nabla_{\sigma}G_{\mu\nu}]=0$, we obtain 
\begin{equation}
\nabla^{\nu}\delta[R_{\mu\nu}]=-3H^2\nabla^{\nu}h_{\mu\nu}+\frac{3}{2}H^2\nabla_{\mu}h+\frac{1}{2}\nabla_{\mu}[\nabla_{\lambda}\nabla^{\lambda}h-\nabla_{\kappa}\nabla_{\lambda}h^{\kappa\lambda}].
\label{P44}
\end{equation}
Finally then, through the use of  (\ref{P44}) and (\ref{P36}) we obtain
\begin{eqnarray}
\delta W^{(2)}_{\mu\nu}&=&\frac{1}{2}[\nabla_{\beta}\nabla^{\beta}+4H^2][\nabla_{\mu}\nabla_{\nu}h
-\nabla_{\mu}\nabla_{\lambda}h^{\lambda}_{\phantom{\lambda}\nu}
-\nabla_{\nu}\nabla_{\lambda}h^{\lambda}_{\phantom{\lambda}\mu}
+\nabla_{\lambda}\nabla^{\lambda}h_{\mu\nu}
-8H^2h_{\mu\nu}+2H^2hg_{\mu\nu}]
\nonumber\\
&+&3H^2\nabla_{\alpha}\nabla^{\alpha}h_{\mu\nu}
+12H^4h_{\mu\nu}
-3H^2\nabla_{\mu}\nabla_{\nu}h
-3H^4g_{\mu\nu}h
\nonumber\\
&+&\frac{1}{2}[g_{\mu\nu}\nabla_{\beta}\nabla^{\beta}-2H^2g_{\mu\nu} -2\nabla_{\mu}\nabla_{\nu}][\nabla_{\lambda}\nabla^{\lambda}h-\nabla_{\kappa}\nabla_{\lambda}h^{\kappa\lambda}]
+\frac{3}{2}g_{\mu\nu}H^2\nabla_{\beta}\nabla^{\beta}h.
\label{P45}
\end{eqnarray}

On making the substitution $h_{\mu\nu}=K_{\mu\nu}+g_{\mu\nu}h/4$, we find that  (\ref{P45}) takes the form  
\begin{eqnarray}
\delta W^{(2)}_{\mu\nu}&=&\frac{1}{2}[\nabla_{\beta}\nabla^{\beta}+4H^2][
-\nabla_{\mu}\nabla_{\lambda}K^{\lambda}_{\phantom{\lambda}\nu}
-\nabla_{\nu}\nabla_{\lambda}K^{\lambda}_{\phantom{\lambda}\mu}
+\nabla_{\lambda}\nabla^{\lambda}K_{\mu\nu}
-8H^2K_{\mu\nu}]
\nonumber\\
&+&3H^2\nabla_{\alpha}\nabla^{\alpha}K_{\mu\nu}
+12H^4K_{\mu\nu}
-\frac{1}{2}[g_{\mu\nu}\nabla_{\beta}\nabla^{\beta}-2H^2g_{\mu\nu} -2\nabla_{\mu}\nabla_{\nu}]\nabla_{\kappa}\nabla_{\lambda}K^{\kappa\lambda}
\nonumber\\
&+&\frac{1}{2}g_{\mu\nu}\nabla_{\alpha}\nabla^{\alpha}\nabla_{\beta}\nabla^{\beta}h
+\frac{1}{4}\nabla_{\alpha}\nabla^{\alpha}\nabla_{\mu}\nabla_{\nu}h
-\frac{3}{4}\nabla_{\mu}\nabla_{\nu}\nabla_{\alpha}\nabla^{\alpha}h
\nonumber\\
&-&2H^2\nabla_{\mu}\nabla_{\nu}h+2H^2g_{\mu\nu}\nabla_{\alpha}\nabla^{\alpha}h.
\label{P46}
\end{eqnarray}
To simplify this expression, we recall a covariant derivative interchange identity that is found (see e.g. \cite{Mannheim2005}) to hold in a background de Sitter geometry for scalars such as $h$, viz.
\begin{equation}
\nabla_{\alpha}\nabla^{\alpha}\nabla_{\mu}\nabla_{\nu}h-\nabla_{\mu}\nabla_{\nu}\nabla_{\alpha}\nabla^{\alpha}h=-2H^2g_{\mu\nu}\nabla_{\alpha}\nabla^{\alpha}h 
+8H^2\nabla_{\mu}\nabla_{\nu}h,
\label{P47}
\end{equation}
with (\ref{P46}) the reducing to
\begin{eqnarray}
\delta W^{(2)}_{\mu\nu}&=&\frac{1}{2}[\nabla_{\beta}\nabla^{\beta}+4H^2][
-\nabla_{\mu}\nabla_{\lambda}K^{\lambda}_{\phantom{\lambda}\nu}
-\nabla_{\nu}\nabla_{\lambda}K^{\lambda}_{\phantom{\lambda}\mu}
+\nabla_{\lambda}\nabla^{\lambda}K_{\mu\nu}
-8H^2K_{\mu\nu}]
\nonumber\\
&+&3H^2\nabla_{\alpha}\nabla^{\alpha}K_{\mu\nu}
+12H^4K_{\mu\nu}
-\frac{1}{2}[g_{\mu\nu}\nabla_{\beta}\nabla^{\beta}-2H^2g_{\mu\nu} -2\nabla_{\mu}\nabla_{\nu}]\nabla_{\kappa}\nabla_{\lambda}K^{\kappa\lambda}
\nonumber\\
&+&\frac{1}{2}g_{\mu\nu}\nabla_{\alpha}\nabla^{\alpha}\nabla_{\beta}\nabla^{\beta}h
-\frac{1}{2}\nabla_{\mu}\nabla_{\nu}\nabla_{\alpha}\nabla^{\alpha}h
+\frac{3}{2}H^2g_{\mu\nu}\nabla_{\alpha}\nabla^{\alpha}h.
\label{P48}
\end{eqnarray}

From the forms for $\delta W^{(1)}_{\mu\nu}$ and $\delta W^{(2)}_{\mu\nu}$ given in (\ref{P39}) and (\ref{P48}) we see that the dependence on the trace $h$ of the fluctuation drops out identically in the conformal combination $\delta W_{\mu\nu}=\delta W^{(2)}_{\mu\nu}-\delta W^{(1)}_{\mu\nu}/3$ required by (\ref{P3}). This decoupling of $h$ from the gravitational fluctuations is precisely as we had anticipated above, and serves as an internal test on our calculation. Only the contribution of $K_{\mu\nu}$ survives, with $\delta W_{\mu\nu}$ finally being given by
\begin{eqnarray}
\delta W_{\mu\nu}&=&\frac{1}{2}[\nabla_{\alpha}\nabla^{\alpha}-4H^2][\nabla_{\beta}\nabla^{\beta}-2H^2]K_{\mu\nu}
\nonumber\\
&-&\frac{1}{2}[\nabla_{\beta}\nabla^{\beta}-4H^2][
\nabla_{\mu}\nabla_{\lambda}K^{\lambda}_{\phantom{\lambda}\nu}
+\nabla_{\nu}\nabla_{\lambda}K^{\lambda}_{\phantom{\lambda}\mu}]
\nonumber\\
&+&\frac{1}{6}[g_{\mu\nu}\nabla_{\alpha}\nabla^{\alpha}+2\nabla_{\mu}\nabla_{\nu}
-6H^2g_{\mu\nu}]\nabla_{\kappa}\nabla_{\lambda}K^{\kappa\lambda},
\label{P49}
\end{eqnarray}
a surprisingly compact form. As constructed, (\ref{P49}) gives the exact first-order perturbative $\delta W_{\mu\nu}$ around a de Sitter background with no choice of coordinate gauge or conformal gauge having been made at all. This then is our main result.

As a second check on our result, we note that as constructed $\delta W_{\mu\nu}$ kinematically obeys the tracelessness condition $g^{\mu\nu}\delta W_{\mu\nu}=0$, just as it has to in a conformal theory.  Finally, as a third check, we recall that in \cite{Mannheim2006} $\delta W_{\mu\nu}$ was evaluated in a flat Cartesian background $\eta_{\mu\nu}$, again with no coordinate gauge or conformal gauge choice having been made, and was found to be of the form
\begin{equation}
\delta W_{\mu\nu}=\frac{1}{2}\Pi^{\rho}_{\phantom{\rho}\mu}
\Pi^{\sigma}_{\phantom{\rho}\nu}K_{\rho \sigma}- \frac{1}{6}\Pi_{\mu \nu} \Pi ^{\rho
\sigma}K_{\rho\sigma},
\label{P50}
\end{equation}
where
\begin{equation}
\Pi_{\mu \nu}
=\eta_{\mu \nu} \partial^{\alpha}\partial_{\alpha}-
\partial_{\mu}\partial_{\nu}.
\label{P51}
\end{equation}
One readily checks that in the flat space limit where one sets $H^2=0$ and replaces covariant derivatives by flat space ones, (\ref{P49}) reduces to (\ref{P50}), just as it should.

\section{Invariance Considerations for de Sitter Fluctuations}
\label{S4}

\subsection{General Considerations}

To analyze the fluctuations around a de Sitter background, we first need to determine the covariant derivatives involved. To this end we recall (see e.g. \cite{Mannheim2005}) that in general in any geometry with metric $g_{MN}(x)$ and Christoffel symbol $\Gamma^{S}_{LR}$, and without regard to any coordinate gauge considerations, the second covariant derivative of a symmetric rank 2 tensor $A_{MN}$ can be written as
\begin{eqnarray}
\nabla_{L}\nabla_{R}A_{MN}&=&
[\partial_{L}\partial_{R}
-\Gamma^{S}_{LR}\partial_{S}]A_{MN}
+[\Gamma^{S}_{LM}\Gamma^{K}_{RN}+
\Gamma^{S}_{LN}\Gamma^{K}_{RM}]A_{KS}
\nonumber \\
&+&
[\Gamma^{S}_{LN}\Gamma^{K}_{RS}+
\Gamma^{S}_{LR}\Gamma^{K}_{SN}
-\partial_{L}\Gamma^{K}_{RN}
-\Gamma^{K}_{RN}\partial_{L}-
\Gamma^{K}_{LN}\partial_{R}]A_{KM}
\nonumber \\
&+&
[\Gamma^{S}_{LM}\Gamma^{K}_{RS}+
\Gamma^{S}_{LR}\Gamma^{K}_{SM}
-\partial_{L}\Gamma^{K}_{RM}
-\Gamma^{K}_{RM}\partial_{L}-
\Gamma^{K}_{LM}\partial_{R}]A_{KN}.
\label{P52}
\end{eqnarray}
While one would need to evaluate (\ref{P52}) directly in the general case, great simplification can be obtained if the $g_{MN}(x)$ geometry is conformal to a flat geometry with Cartesian metric $\eta_{MN}$, and the line element is given by (\ref{P6}) with an appropriate conformal factor $\Omega^2(x)$. In this case the $g_{MN}(x)$ geometry Christoffel symbols evaluate to
\begin{equation}
\Gamma^{L}_{MN}=\Omega^{-1}(x)[\delta^{L}_{M}\partial_{N}
+\delta^{L}_{N}\partial_{M}-\eta^{LR}\eta_{MN}\partial_{R}]\Omega(x).
\label{P53}
\end{equation}
Following some algebra the general $\nabla_{L}\nabla^{L}A_{MN}$ is then found to evaluate to
\begin{eqnarray}
g^{LR}\nabla_{L}\nabla_{R}A_{MN}&=&
\eta^{LR}\Omega^{-2}\partial_{L}\partial_{R}A_{MN}
+2\Omega^{-4}\partial_{M}\Omega\partial_{N}\Omega \eta^{TQ}A_{TQ}
-2\eta^{LR}\Omega^{-3}\partial_{L}\partial_{R}\Omega A_{MN}
\nonumber \\
&-&2\eta^{LR}\Omega^{-3}\partial_{R}\Omega \partial_{L}A_{MN}
+2\Omega^{-4}\eta_{MN}\eta^{TX}\partial_{X}\Omega\eta^{QY} \partial_{Y}\Omega A_{TQ}
\nonumber \\
&-&4\Omega^{-4}\partial_{M}\Omega \eta^{TX}\partial_{X}\Omega A_{TN}
-4\Omega^{-4}\partial_{N}\Omega \eta^{TX}\partial_{X}\Omega A_{TM}
\nonumber \\
&+&2\eta^{KQ}\Omega^{-3}\partial_{Q}\Omega \partial_{N}A_{KM}
+2\eta^{KQ}\Omega^{-3}\partial_{Q}\Omega \partial_{M}A_{KN}
\nonumber\\
&-&2\eta^{LR}\Omega^{-3}\partial_{N}\Omega \partial_{L}A_{RM}
-2\eta^{LR}\Omega^{-3}\partial_{M}\Omega \partial_{L}A_{RN}.
\label{P54}
\end{eqnarray}
In (\ref{P54}) everything is now conveniently expressed in terms of just the one function $\Omega(x)$ alone.

By the same token, the covariant derivative of $A_{MN}$ is given by
\begin{equation}
\nabla_{P}A^{P}_{\phantom{P}M}=\Omega^{-2}\eta^{LP}\partial_{P}A_{LM}
+2\eta^{LR}\Omega^{-3}\partial_{L}\Omega A_{RM}
-\Omega^{-3}\eta^{LR}\partial_{M}\Omega A_{LR}.
\label{P55}
\end{equation}
Given (\ref{P55}), we can rewrite (\ref{P54}) as 
\begin{eqnarray}
g^{LR}\nabla_{L}\nabla_{R}A_{MN}&=&
\eta^{LR}\Omega^{-2}\partial_{L}\partial_{R}A_{MN}
-2\Omega^{-4}\partial_{M}\Omega\partial_{N}\Omega \eta^{TQ}A_{TQ}
-2\eta^{LR}\Omega^{-3}\partial_{L}\partial_{R}\Omega A_{MN}
\nonumber \\
&-&2\eta^{LR}\Omega^{-3}\partial_{R}\Omega \partial_{L}A_{MN}
+2\Omega^{-4}\eta_{MN}\eta^{TX}\partial_{X}\Omega\eta^{QY} \partial_{Y}\Omega A_{TQ}
\nonumber \\
&+&2\eta^{KQ}\Omega^{-3}\partial_{Q}\Omega \partial_{N}A_{KM}
+2\eta^{KQ}\Omega^{-3}\partial_{Q}\Omega \partial_{M}A_{KN}
\nonumber\\
&-&2\Omega^{-1}\partial_{N}\Omega \nabla_{L}A^{L}_{\phantom{L}M}
-2\Omega^{-1}\partial_{M}\Omega \nabla_{L}A^{L}_{\phantom{L}N},
\label{P56}
\end{eqnarray}
where the $\nabla_{L}A^{L}_{\phantom{L}M}$ and $\nabla_{L}A^{L}_{\phantom{L}N}$ terms refer to covariant derivatives in the $g_{MN}$ geometry. In addition, we note that for an $A_{MN}$ that is  traceless, and for an $\Omega$ that only depends on the time coordinate $t$, the quantity $g^{LR}\nabla_{L}\nabla_{R}A_{MN}$ in (\ref{P56}) would simplify a great deal  if $A_{MN}$ were to obey the synchronous condition $A_{0M}=0$ for all four values of $M$, as it  would then reduce to  
\begin{eqnarray}
g^{LR}\nabla_{L}\nabla_{R}A_{MN}&=&
\eta^{LR}\Omega^{-2}\partial_{L}\partial_{R}A_{MN}
-2\eta^{LR}\Omega^{-3}\partial_{L}\partial_{R}\Omega A_{MN}
-2\eta^{LR}\Omega^{-3}\partial_{R}\Omega \partial_{L}A_{MN}
\nonumber\\
&-&2\Omega^{-1}\partial_{N}\Omega \nabla_{L}A^{L}_{\phantom{L}M}
-2\Omega^{-1}\partial_{M}\Omega \nabla_{L}A^{L}_{\phantom{L}N}.
\label{P57}
\end{eqnarray}
In (\ref{P57}) it is only the $\nabla_{L}A^{L}_{\phantom{L}M}$ terms (now equal to $\Omega^{-2}\eta^{LP}\partial_{P}A_{LM}$) that then prevent $g^{LR}\nabla_{L}\nabla_{R}A_{MN}$ from being diagonal in its $(M, N)$ indices. Moreover, if a traceless, synchronous $A_{MN}$ is  transverse, then in an $\Omega(t)$ geometry $A_{MN}$ would only have two independent components, with its six spatial components obeying the four spatially transverse-traceless conditions $\eta^{ij}\partial_{j}A_{ik}=0$, $\eta^{ij}A_{ij}=0$. 

\subsection{General Structure of $\delta \bar{W}_{\mu\nu}$ in the de Sitter Case}

At this point it is convenient  to now revert back to our earlier notation and let $\bar{g}^{(0)}_{\mu\nu}(x)$ denote the conformal to flat background de Sitter metric and $\bar{K}_{\mu\nu}(x)$ denote the fluctuation around it. Moreover, at this point it is also convenient to take advantage of the coordinate invariance of the theory and impose the transverse gauge condition 
\begin{equation}
\bar{\nabla}_{\nu}\bar{K}^{\mu\nu}=0.
\label{P58}
\end{equation}
In this gauge we find that the first-order fluctuation function $\delta \bar{W}_{\mu\nu}$ given in (\ref{P49}) then reduces to the very simple form
\begin{equation}
\delta \bar{W}_{\mu\nu}=\frac{1}{2}[\bar{\nabla}_{\alpha}\bar{\nabla}^{\alpha}-4H^2][\bar{\nabla}_{\beta}\bar{\nabla}^{\beta}-2H^2]\bar{K}_{\mu\nu}.
\label{P59}
\end{equation}

For any $\bar{A}_{\mu\nu}$  that obeys $\bar{\nabla}_{\nu}\bar{A}^{\mu\nu}=0$ in the barred notation, (\ref{P56}) takes the form 
\begin{eqnarray}
\bar{g}_{(0)}^{\lambda\rho}\bar{\nabla}_{\lambda}\bar{\nabla}_{\rho}\bar{A}_{\mu\nu}&=&
\eta^{\lambda\rho}\Omega^{-2}\partial_{\lambda}\partial_{\rho}\bar{A}_{\mu\nu}
-2\Omega^{-4}\partial_{\mu}\Omega\partial_{\nu}\Omega \eta^{\tau\sigma}\bar{A}_{\tau\sigma}
-2\eta^{\lambda\rho}\Omega^{-3}\partial_{\lambda}\partial_{\rho}\Omega \bar{A}_{\mu\nu}
\nonumber \\
&-&2\eta^{\lambda\rho}\Omega^{-3}\partial_{\lambda}\Omega \partial_{\rho}\bar{A}_{\mu\nu}
+2\Omega^{-4}\eta_{\mu\nu}\eta^{\tau\alpha}\partial_{\alpha}\Omega\eta^{\sigma\beta} \partial_{\beta}\Omega \bar{A}_{\tau\sigma}
\nonumber \\
&+&2\eta^{\kappa\sigma}\Omega^{-3}\partial_{\sigma}\Omega \partial_{\nu}\bar{A}_{\kappa\mu}
+2\eta^{\kappa\sigma}\Omega^{-3}\partial_{\sigma}\Omega \partial_{\mu}\bar{A}_{\kappa\nu}.
\label{P60}
\end{eqnarray}
For the purposes of evaluating $\delta \bar{W}_{\mu\nu}$, we can set $\bar{A}_{\mu\nu}=\bar{K}_{\mu\nu}$ in (\ref{P60}) in order to determine $\bar{\nabla}_{\beta}\bar{\nabla}^{\beta}\bar{K}_{\mu\nu}$. Since $\bar{\nabla}_{\beta}\bar{\nabla}^{\beta}\bar{K}_{\mu\nu}$  is itself a rank 2 tensor we would like to be able to set $\bar{A}_{\mu\nu}=\bar{\nabla}_{\beta}\bar{\nabla}^{\beta}\bar{K}_{\mu\nu}$ in (\ref{P60}) as well so as to determine $\bar{\nabla}_{\alpha}\bar{\nabla}^{\alpha}\bar{\nabla}_{\beta}\bar{\nabla}^{\beta}\bar{K}_{\mu\nu}$. However, in order to use (\ref{P60}) with its transverse $\bar{\nabla}_{\nu}\bar{A}^{\mu\nu}=0$, we note that while we have taken $\bar{K}_{\mu\nu}$ to be transverse,  it does not immediately follow that $\bar{\nabla}_{\beta}\bar{\nabla}^{\beta}\bar{K}_{\mu\nu}$ will be transverse too. However on  using the covariant derivative relations
\begin{equation}
\nabla_{P}\nabla_{K}\nabla_{N}A_{LM}-\nabla_{K}\nabla_{P}\nabla_{N}A_{LM}=R_{LSKP}\nabla_{N}A^{S}_{\phantom{S}M}
+R_{MSKP}\nabla_{N}A^{S}_{\phantom{S}L}
+R_{NSKP}\nabla^{S}A_{LM},
\label{P61}
\end{equation}
\begin{equation}
\nabla_{P}\nabla_{N}A_{LM}-\nabla_{N}\nabla_{P}A_{LM}=
R_{LSNP}A^{S}_{\phantom{S}M}
+R_{MSNP}A^{S}_{\phantom{S}L}
\label{P62}
\end{equation}
that hold in any geometry, for a de Sitter geometry we obtain
\begin{equation}
\nabla_{P}\nabla_{K}\nabla^{K}A^{P}_{\phantom{P}M}
=\nabla_{K}\nabla^{K}\nabla_{P}A^{P}_{\phantom{P}M}
+5H^2\nabla_{P}A^{P}_{\phantom{P}M}
-2H^2\nabla_{M}A^{P}_{\phantom{P}P}.
\label{P63}
\end{equation}
Consequently, if $A_{MN}$ is transverse and traceless, then in a de Sitter geometry so is $\nabla_{K}\nabla^{K}A_{MN}$. 

At this point by taking $\bar{A}_{\mu\nu}=\bar{K}_{\mu\nu}$ and then $\bar{A}_{\mu\nu}=\bar{\nabla}_{\beta}\bar{\nabla}^{\beta}\bar{K}_{\mu\nu}$, we can now evaluate $\delta \bar{W}_{\mu\nu}$ as given in (\ref{P59}) using the general expression 
\begin{eqnarray}
\bar{g}_{(0)}^{\lambda\rho}\bar{\nabla}_{\lambda}\bar{\nabla}_{\rho}\bar{A}_{\mu\nu}&=&
\eta^{\lambda\rho}\Omega^{-2}\partial_{\lambda}\partial_{\rho}\bar{A}_{\mu\nu}
-2\eta^{\lambda\rho}\Omega^{-3}\partial_{\lambda}\partial_{\rho}\Omega \bar{A}_{\mu\nu}
-2\eta^{\lambda\rho}\Omega^{-3}\partial_{\lambda}\Omega \partial_{\rho}\bar{A}_{\mu\nu}
\nonumber\\
&+&2\Omega^{-4}\eta_{\mu\nu}\eta^{\tau\alpha}\partial_{\alpha}\Omega\eta^{\sigma\beta} \partial_{\beta}\Omega \bar{A}_{\tau\sigma}
+2\eta^{\kappa\sigma}\Omega^{-3}\partial_{\sigma}\Omega \partial_{\nu}\bar{A}_{\kappa\mu}
+2\eta^{\kappa\sigma}\Omega^{-3}\partial_{\sigma}\Omega \partial_{\mu}\bar{A}_{\kappa\nu}
\label{P64}
\end{eqnarray}
that holds for any transverse-traceless $\bar{A}_{\mu\nu}$. This is far as we can go in reducing the problem without actually imposing the equation of motion $\delta \bar{W}_{\mu\nu}=0$ that the fluctuations are to obey.

\subsection{Solution to the Fluctuation Equations}

When we constrain $\delta \bar{W}_{\mu\nu}$ by requiring it to obey an equation of motion of the form $\delta \bar{W}_{\mu\nu}=0$, we then have the opportunity to utilize any additional gauge symmetries or structure that solutions to the fluctuation equations might possess. And in fact if we seek solutions that obey a synchronous condition, we find that we can consistently satisfy all constraints simultaneously. Specifically, for any transverse-traceless $\bar{A}_{\mu\nu}$ that obeys $\bar{A}_{0 \mu}=0$, (\ref{P64}) reduces to 
\begin{equation}
\bar{g}_{(0)}^{\lambda\rho}\bar{\nabla}_{\lambda}\bar{\nabla}_{\rho}\bar{A}_{\mu\nu}=
\eta^{\lambda\rho}\Omega^{-2}\partial_{\lambda}\partial_{\rho}\bar{A}_{\mu\nu}
-2\eta^{\lambda\rho}\Omega^{-3}\partial_{\lambda}\partial_{\rho}\Omega \bar{A}_{\mu\nu}
-2\eta^{\lambda\rho}\Omega^{-3}\partial_{\lambda}\Omega \partial_{\rho}\bar{A}_{\mu\nu},
\label{P65}
\end{equation}
to now be diagonal in its $(\mu,\nu)$ indices. Since (\ref{P65}) now is diagonal, we can consistently set $\bar{g}_{(0)}^{\lambda\rho}\bar{\nabla}_{\lambda}\bar{\nabla}_{\rho}\bar{A}_{0\mu}=0$ for the four $\bar{A}_{0\mu}$ components. Moreover, since (\ref{P65}) is diagonal in its indices, it will remain so whether we take $\bar{A}_{\mu\nu}$ to be equal to $\bar{K}_{\mu\nu}$ or $\bar{\nabla}_{\beta}\bar{\nabla}^{\beta}\bar{K}_{\mu\nu}$. Consequently, all nine components of $\bar{K}_{\mu\nu}$ will obey 
\begin{eqnarray}
\bar{\nabla}_{\alpha}\bar{\nabla}^{\alpha}\bar{\nabla}_{\lambda}\bar{\nabla}^{\lambda}\bar{K}_{\mu\nu}
&=&[\eta^{\alpha\beta}\Omega^{-2}\partial_{\alpha}\partial_{\beta}
-2\eta^{\alpha\beta}\Omega^{-3}\partial_{\alpha}\partial_{\beta}\Omega 
-2\eta^{\alpha\beta}\Omega^{-3}\partial_{\alpha}\Omega \partial_{\beta}]
\nonumber\\
&\times&
[\eta^{\lambda\rho}\Omega^{-2}\partial_{\lambda}\partial_{\rho}
-2\eta^{\lambda\rho}\Omega^{-3}\partial_{\lambda}\partial_{\rho}\Omega 
-2\eta^{\lambda\rho}\Omega^{-3}\partial_{\lambda}\Omega \partial_{\rho}]\bar{K}_{\mu\nu},
\label{P66}
\end{eqnarray}
some components doing so trivially and some doing so non-trivially. 

While there would appear to be nine conditions, viz. $g_{(0)}^{\mu\nu}\bar{K}_{\mu\nu}=0$, $\bar{\nabla}_{\nu}\bar{K}^{\mu\nu}=0$ and $\bar{K}^{0\mu}=0$, there is some overlap, with two independent components surviving, just as one would expect of a massless rank 2 tensor theory. Because of the synchronous condition and because the transverse condition reduces to  $\eta^{\lambda\nu}\partial_{\nu}\bar{K}_{\lambda\mu}=0$, the surviving components have purely spatial indices and obey the four spatially transverse-traceless conditions $\eta^{ij}\partial_{j}\bar{K}_{ik}=0$, $\eta^{ij}\bar{K}_{ij}=0$, to give precisely two independent components. In the cosmological literature these modes are known as tensor modes.

To solve (\ref{P66}) we need to determine the explicit form for $\Omega$ in the de Sitter case. Thus if we start with the de Sitter metric in the comoving form
\begin{equation}
ds^2=-dt^2+e^{2Ht}[dx^2+dy^2+dz^2],
\label{P67}
\end{equation}
and make the coordinate transformation
\begin{equation}
d\tau=\frac{dt}{e^{Ht}},\qquad \tau=-\frac{e^{-Ht}}{H},
\label{P68}
\end{equation}
we find that the line element takes the conformal to flat form 
\begin{equation}
ds^2=\frac{1}{\tau^2H^2}[-d\tau^2+dx^2+dy^2+dz^2].
\label{P69}
\end{equation}
We thus identify $\Omega=1/\tau H$.

With this form for $\Omega(\tau)$ we find that $\bar{\nabla}_{\beta}\bar{\nabla}^{\beta}\bar{K}_{\mu\nu}$ takes the form
\begin{equation}
\bar{\nabla}_{\beta}\bar{\nabla}^{\beta}\bar{K}_{\mu\nu}=\tau^2H^2\left[-\frac{\partial^2}{\partial \tau^2}+\eta^{ij}\partial_{i}\partial_{j}-\frac{2}{\tau}\frac{\partial}{\partial \tau}+\frac{4}{\tau^2}\right]\bar{K}_{\mu\nu}.
\label{P70}
\end{equation}
In a 3-momentum eigenstate $\bar{q}$ we can set $\bar{K}_{\mu\nu}(\tau,\bar{x})=\bar{K}_{\mu\nu}(\tau)e^{i\bar{q}\cdot\bar{x}}$, with (\ref{P59}) then reducing to
\begin{eqnarray}
&&[\bar{\nabla}_{\alpha}\bar{\nabla}^{\alpha}-4H^2][\bar{\nabla}_{\beta}\bar{\nabla}^{\beta}-2H^2][\bar{K}_{\mu\nu}(\tau)e^{i\bar{q}\cdot\bar{x}}]
\nonumber\\
&&=\tau^4H^4\left[\frac{\partial^4}{\partial \tau^4}
+\frac{8}{\tau}\frac{\partial^3}{\partial \tau^3}
+\frac{12}{\tau^2}\frac{\partial^2}{\partial \tau^2}
+2q^2\left(\frac{\partial^2}{\partial \tau^2}
+\frac{4}{\tau}\frac{\partial}{\partial \tau}
+\frac{2}{\tau^2}\right)+q^4\right]\left[\bar{K}_{\mu\nu}(\tau)e^{i\bar{q}\cdot\bar{x}}\right].
\label{P71}
\end{eqnarray}
Finally, the expression that appears in (\ref{P71}) can be factored, with the fluctuation equation being found to reduce to the remarkably simple
\begin{equation}
\tau^2H^2\left[-\frac{\partial^2}{\partial \tau^2}-q^2\right]\left[-\frac{\partial^2}{\partial \tau^2}-q^2\right]
\left[\tau^2H^2\bar{K}_{\mu\nu}(\tau)e^{i\bar{q}\cdot\bar{x}}\right]=0.
\label{P72}
\end{equation}
With (\ref{P72}) we achieve our primary objective since we can now determine $\bar{K}_{\mu\nu}$ in a closed form.

Since according to (\ref{P18}) the relation between the barred and unbarred fluctuations is given by $\bar{K}_{\mu\nu}(x)=\Omega^2(x)K_{\mu\nu}(x)$, we see that we can identify $\tau^2H^2\bar{K}_{\mu\nu}(x)$ with $K_{\mu\nu}(x)$, the fluctuation associated with a flat background metric.  However, in a flat background in the transverse gauge $\partial_{\nu}K^{\mu\nu}=0$ (a gauge that is conformally equivalent to the $\bar{\nabla}_{\nu}\bar{K}^{\mu\nu}=0$ gauge for synchronous modes), we find that on setting to zero the flat space background  $\delta W_{\mu\nu}$ given in (\ref{P50}) we obtain none other than 
\begin{equation}
\left[-\frac{\partial^2}{\partial \tau^2}-q^2\right]\left[-\frac{\partial^2}{\partial \tau^2}-q^2\right]
\left[K_{\mu\nu}(\tau)e^{i\bar{q}\cdot\bar{x}}\right]=0.
\label{P73}
\end{equation}
As we had  noted above, this is just as is to be expected on general grounds. Thus, for the de Sitter case we explicitly confirm that the fluctuations associated with a conformal to flat background can be constructed via a conformal transformation on the fluctuations associated with a flat background, even though the perturbed geometry is not itself conformal to flat.

In \cite{Mannheim2009} tensor mode solutions to (\ref{P73}) were explicitly obtained, with there being two positive frequency solutions of the form of
\begin{equation}
K_{\mu\nu}(\tau,\bar{x})=A_{\mu\nu}e^{i\bar{q}\cdot\bar{x}-iq\tau}+B_{\mu\nu}\tau e^{i\bar{q}\cdot\bar{x}-iq\tau},
\label{P74}
\end{equation}
where $A_{\mu\nu}$ and $B_{\mu\nu}$ are polarization tensors. Of the two solutions the $A_{\mu\nu}$ mode satisfies the second-order $[-\partial_{\tau}^2-q^2]A_{\mu\nu}=0$, while the $B_{\mu\nu}$ mode is an intrinsically fourth-order mode that is not encountered in the second-order theory at all \cite{footnote000}. (The action of $[-\partial_{\tau}^2-q^2]$ on the $B_{\mu\nu}$ term yields the plane wave $e^{i\bar{q}\cdot\bar{x}-i\tau q}$ so that $[-\partial_\tau^2-q^2]^2B_{\mu\nu}=0$.) Finally, given (\ref{P74}), we see that the positive frequency solutions (and analogously the negative frequency solutions) to (\ref{P71}) are given by
\begin{equation}
\bar{K}_{\mu\nu}(\tau,\bar{x})=\frac{1}{\tau^2H^2}A_{\mu\nu}e^{i\bar{q}\cdot\bar{x}-iq\tau}+\frac{1}{\tau H^2}B_{\mu\nu}e^{i\bar{q}\cdot\bar{x}-iq\tau}. 
\label{P75}
\end{equation}

Up to coordinate transformations the solution given in (\ref{P75}) is both exact and the most general solution for first-order perturbative tensor fluctuations around a de Sitter background in the conformal gravity theory. Finally, since coordinate transformations involving the time coordinate do not affect $\bar{K}_{\mu\nu}(\tau,\bar{x})$ modes that are synchronous, in comoving coordinates the solutions in (\ref{P75}) take the form
\begin{equation}
\bar{K}_{\mu\nu}(t,\bar{x})=e^{2Ht}A_{\mu\nu}\exp[i\bar{q}\cdot\bar{x}+ie^{-Ht}q/H]-\frac{e^{Ht}}{H}B_{\mu\nu}\exp[i\bar{q}\cdot\bar{x}+ie^{-Ht}q/H].
\label{P76}
\end{equation}

\subsection{Comparison with the Second-Order Case}

To analyze fluctuations around a de Sitter background in the standard second-order theory, in (\ref{P5}) we set $\bar{T}^{(0)}_{\mu\nu}=-(3H^2/8\pi G) \bar{g}^{(0)}_{\mu\nu}$, with the background equations then being of the form
\begin{equation}
\bar{R}^{(0)}_{\mu\nu} -\frac{1}{2}\bar{g}^{(0)}_{\mu\nu}\bar{g}_{(0)}^{\alpha\beta}\bar{R}^{(0)}_{\alpha\beta}-3H^2\bar{g}^{(0)}_{\mu\nu}=0.
\label{P77}
\end{equation}
Consequently, around the de Sitter background the fluctuation equations take the form
\begin{equation}
[\bar{\nabla}_{\alpha}\bar{\nabla}^{\alpha}-2H^2]\bar{h}_{\mu\nu}
+\bar{\nabla}_{\mu}\bar{\nabla}_{\nu}\bar{h}
-\bar{\nabla}_{\mu}\bar{\nabla}_{\lambda}\bar{h}^{\lambda}_{\phantom{\lambda}\nu}
-\bar{\nabla}_{\nu}\bar{\nabla}_{\lambda}\bar{h}^{\lambda}_{\phantom{\lambda}\mu}
-\bar{g}^{(0)}_{\mu\nu}[\bar{\nabla}_{\lambda}\bar{\nabla}^{\lambda}\bar{h}-\bar{\nabla}_{\kappa}\bar{\nabla}_{\lambda}\bar{h}^{\kappa\lambda}+H^2\bar{h}]=0.
\label{P78}
\end{equation}
On restricting to fluctuations that are transverse and traceless, we see that (\ref{P78}) reduces to
\begin{equation}
[\bar{\nabla}_{\alpha}\bar{\nabla}^{\alpha}-2H^2]\bar{h}_{\mu\nu}=0.
\label{P79}
\end{equation}
Comparing with (\ref{P59}), we thus see that the transverse-traceless fluctuations that occur in  standard gravity are also solutions to the conformal theory. A similar outcome was found for fluctuations around a flat background, with the $A_{\mu\nu}e^{i\bar{q}\cdot\bar{x}-iq\tau}$ type modes given in (\ref{P74}) also occurring in the second-order case. Since the conformal theory is fourth order, it also possesses fluctuation solutions that do not occur in the standard case, viz. the solutions to $[\bar{\nabla}_{\alpha}\bar{\nabla}^{\alpha}-4H^2]\bar{K}_{\mu\nu}=0$ in the de Sitter background case, and the  $B_{\mu\nu}\tau e^{i\bar{q}\cdot\bar{x}-iq\tau}$ type modes given in (\ref{P74})  in the flat background case.

However, the $[\bar{\nabla}_{\alpha}\bar{\nabla}^{\alpha}-2H^2]\bar{h}_{\mu\nu}=0$ and $[\bar{\nabla}_{\alpha}\bar{\nabla}^{\alpha}-2H^2]\bar{K}_{\mu\nu}=0$ de Sitter fluctuation solutions in the second- and fourth-order cases are not in one to one correspondence with each other, since a conformal transformation on the flat  $A_{\mu\nu}e^{i\bar{q}\cdot\bar{x}-iq\tau}$ type solution cannot  bring one to the second-order $[\bar{\nabla}_{\alpha}\bar{\nabla}^{\alpha}-2H^2]\bar{h}_{\mu\nu}=0$ solution since the second-order theory is not conformal invariant. Rather, a conformal transformation on $A_{\mu\nu}e^{i\bar{q}\cdot\bar{x}-iq\tau}$ must bring one to some linear combination of the solutions to the $[\bar{\nabla}_{\alpha}\bar{\nabla}^{\alpha}-2H^2]\bar{K}_{\mu\nu}=0$ and $[\bar{\nabla}_{\alpha}\bar{\nabla}^{\alpha}-4H^2]\bar{K}_{\mu\nu}=0$ equations.

To see what specifically does happen when we make a conformal transformation, we use (\ref{P70}) to write out the relevant equations for synchronous modes with 3-momentum $\bar{q}$, viz.
\begin{equation}
[\bar{\nabla}_{\alpha}\bar{\nabla}^{\alpha}-2H^2]\bar{A}_{\mu\nu}
=\tau^2H^2\left[-\frac{\partial^2}{\partial \tau^2}-q^2-\frac{2}{\tau}\frac{\partial}{\partial \tau}+\frac{2}{\tau^2}\right]\bar{A}_{\mu\nu},
\label{P80}
\end{equation}
\begin{equation}
[\bar{\nabla}_{\alpha}\bar{\nabla}^{\alpha}-4H^2]\bar{A}_{\mu\nu}
=\tau^2H^2\left[-\frac{\partial^2}{\partial \tau^2}-q^2-\frac{2}{\tau}\frac{\partial}{\partial \tau}\right]\bar{A}_{\mu\nu},
\label{P81}
\end{equation}
where $\bar{A}_{\mu\nu}$ denotes $\bar{h}_{\mu\nu}$ or $\bar{K}_{\mu\nu}$ as appropriate. The positive frequency solution to (\ref{P80}) with linear 3-momentum $\bar{q}$ is of the form
\begin{equation}
\bar{A}_{\mu\nu}(\tau,\bar{x})=
\alpha_{\mu\nu}e^{i\bar{q}\cdot\bar{x}-iq\tau}\left[\frac{1}{\tau^2}+\frac{iq}{\tau}\right],
\label{P82}
\end{equation}
where $\alpha_{\mu\nu}$ is a polarization tensor. Similarly, the positive frequency solution to (\ref{P81}) is of the form
\begin{equation}
\bar{A}_{\mu\nu}(\tau,\bar{x})=
\beta_{\mu\nu}\frac{e^{i\bar{q}\cdot\bar{x}-iq\tau}}{\tau}
\label{P83}
\end{equation}
with polarization tensor $\beta_{\mu\nu}$.
Thus while the fourth-order solution $(1/\tau^2H^2)A_{\mu\nu}e^{i\bar{q}\cdot\bar{x}-iq\tau}$  given in (\ref{P75}) is obtained via a conformal transformation on the $A_{\mu\nu}e^{i\bar{q}\cdot\bar{x}-iq\tau}$ solution given in (\ref{P75}), this solution is not a solution to the second-order theory (i.e. it does not obey the second-order $[\bar{\nabla}_{\alpha}\bar{\nabla}^{\alpha}-2H^2]A_{\mu\nu}=0$), even though the $A_{\mu\nu}e^{i\bar{q}\cdot\bar{x}-iq\tau}$ solution does obey  the second-order equation $\partial_{\alpha}\partial^{\alpha}A_{\mu\nu}=0$.

Finally, writing the second-order theory de Sitter background fluctuation solution given in (\ref{P82}) in the  comoving coordinate system, we obtain 
\begin{equation}
\bar{h}_{\mu\nu}(\tau,\bar{x})=\alpha_{\mu\nu}\exp[i\bar{q}\cdot\bar{x}+ie^{-Ht}q/H]
[H^2e^{2Ht}-iqHe^{Ht}].
\label{P84}
\end{equation}
Comparing with (\ref{P76}), we see that for tensor fluctuations around a de Sitter background the leading $e^{2Ht}$ behavior in comoving time is the same for both the second- and fourth-order theories, and both would thus give rise to very rapid growth (and for the particular conformal gravity fluctuation mode with $B_{\mu\nu}=iqA_{\mu\nu}$ the two theories would even give the exact same growth).

\section{Fluctuations Around Robertson-Walker Backgrounds}
\label{S5}

\subsection{Cartesian Coordinate Considerations}

In comoving coordinates the line element for each of the three Robertson-Walker geometries takes the form
\begin{equation}
ds^2=-dt^2+a^2(t)\left[\frac{dr^2}{1-Kr^2} +r^2d\theta^2+r^2\sin^2\theta d\phi^2\right],
\label{P85}
\end{equation}
where the 3-curvature $K$ can be positive, zero or negative. For each of these geometries we can make a coordinate transformation on the time coordinate of the form
\begin{equation}
d\tau=\frac{dt}{a(t)}
\label{P86}
\end{equation}
and bring the line element to the form
\begin{equation}
ds^2=a^2(t)\left[-d\tau^2+\frac{dr^2}{1-Kr^2} +r^2d\theta^2+r^2\sin^2\theta d\phi^2\right].
\label{P87}
\end{equation}

For the spatially flat $K=0$ RW geometry we can make a further coordinate transformation from polar coordinates to Cartesian ones, and bring the line element to the form 
\begin{equation}
ds^2=a^2(t)\left[-d\tau^2+dx^2+dy^2+dz^2\right].
\label{P88}
\end{equation}
Since this line element is now in the form given in (\ref{P6}) and since the conformal factor $\Omega$ only depends on the time coordinate, our analysis of fluctuations around a  de Sitter geometry carries over completely, with an $a(t)$ that can now be of a more general form than the $a(t)=e^{Ht}=-1/\tau H$ form used in the de Sitter case. Thus we can again use (\ref{P74}) for $K_{\mu\nu}$ and can again use (\ref{P18}) to construct $\bar{K}_{\mu\nu}$, and can again take the fluctuations to be transverse, traceless and synchronous, this time with respect to the metric given in (\ref{P88}). Consequently, around the spatially flat background given in (\ref{P88}) the positive frequency conformal gravity fluctuations that obey $\eta^{ij}\partial_{j}\bar{K}_{ik}=0$, $\eta^{ij}\bar{K}_{ij}=0$, $\bar{K}_{00}=0$, $\bar{K}_{0i}=0$ and have 3-momentum $\bar{q}$ are given in closed form as
\begin{equation}
\bar{K}_{\mu\nu}(\tau,\bar{x})=a^2(t)A_{\mu\nu}e^{i\bar{q}\cdot\bar{x}-iq\tau}+a^2(t)B_{\mu\nu}\tau e^{i\bar{q}\cdot\bar{x}-iq\tau}. 
\label{P89}
\end{equation}
In this expression the coordinate $\tau$ can then explicitly be transformed into the comoving $t$ once the dependence of $a(t)$ on $t$ is specified.

To bring the RW geometries with non-zero $K$ to a conformal to flat form requires coordinate transformations that involve both $\tau$ and $r$. For the $K>0$ case first, it is convenient to set $K=1/L^2$, and introduce $\sin \chi=r/L$, with (\ref{P87}) then taking the form
\begin{equation}
ds^2=L^2a^2(t)\left[-dp^2+d\chi^2 +\sin^2\chi d\theta^2+\sin^2\chi \sin^2\theta d\phi^2\right],
\label{P90}
\end{equation}
where $p=\tau/L$. Next we introduce
\begin{equation}
p^{\prime}+r^{\prime}=\tan[(p+\chi)/2],\qquad p^{\prime}-r^{\prime}=\tan[(p-\chi)/2],
\label{P91}
\end{equation}
so that
\begin{eqnarray}
-dp^{\prime 2}+dr^{\prime 2}&=&\frac{1}{4}[-dp^2+d\chi^2]\sec^2[(p+\chi)/2]\sec^2[(p-\chi)/2],
\nonumber\\
\sin\chi&=&\sin[(p+\chi)/2]\cos[(p-\chi)/2]-\sin[(p-\chi)/2]\cos[(p+\chi)/2]
\nonumber\\
&=&\left[\tan[(p+\chi)/2]-\tan[(p-\chi)/2]\right]\cos[(p+\chi)/2]\cos[(p-\chi)/2]
\nonumber\\
&=&2r^{\prime}\cos[(p+\chi)/2]\cos[(p-\chi)/2],
\nonumber\\
\sec^2[(p+\chi)/2]\sec^2[(p-\chi)/2]&=&[1+(p^{\prime}+r^{\prime})^2][1+(p^{\prime}-r^{\prime})^2].
\label{P92}
\end{eqnarray}
With these transformations the line element then takes the conformal to flat form
\begin{equation}
ds^2=\frac{4L^2a^2(t)}{[1+(p^{\prime}+r^{\prime})^2][1+(p^{\prime}-r^{\prime})^2]}\left[-dp^{\prime 2}+dr^{\prime 2} +r^{\prime 2}d\theta^2+r^{\prime 2} \sin^2\theta d\phi^2\right].
\label{P93}
\end{equation}
Now while we could transform the spatial sector  of (\ref{P93}) from polar coordinates to Cartesian ones via $x^{\prime}=r^{\prime}\sin\theta\cos\phi$, $y^{\prime}=r^{\prime}\sin\theta\sin\phi$, $z^{\prime}=r^{\prime}\cos\theta$ and thus bring the line element into the form given in (\ref{P6}), the resulting conformal factor would depend on $p^{\prime}$ and $(x^{\prime 2}+ y^{\prime 2}+z^{\prime 2})^{1/2}$, and thus not straightforwardly permit any of the synchronous mode type simplifications we found when the conformal factor only depended on the time coordinate. We shall thus find it more convenient to continue to work in polar coordinates and provide the polar coordinate analysis below.

For the $K<0$ case, it is convenient to set $K=-1/L^2$, and introduce ${\rm sinh} \chi=r/L$, with (\ref{P87}) then taking the form
\begin{equation}
ds^2=L^2a^2(t)\left[-dp^2+d\chi^2 +{\rm sinh}^2\chi d\theta^2+{\rm sinh}^2\chi \sin^2\theta d\phi^2\right],
\label{P94}
\end{equation}
where $p=\tau/L$. Next we introduce
\begin{equation}
p^{\prime}+r^{\prime}=\tanh[(p+\chi)/2],\qquad p^{\prime}-r^{\prime}=\tanh[(p-\chi)/2],
\label{P95}
\end{equation}
so that
\begin{eqnarray}
-dp^{\prime 2}+dr^{\prime 2}&=&\frac{1}{4}[-dp^2+d\chi^2]{\rm sech}^2[(p+\chi)/2]{\rm sech}^2[(p-\chi)/2],
\nonumber\\
{\rm sinh}\chi&=&2r^{\prime}{\rm cosh}[(p+\chi)/2]{\rm cosh}[(p-\chi)/2],
\nonumber\\
{\rm sech}^2[(p+\chi)/2]{\rm sech}^2[(p-\chi)/2]&=&[1-(p^{\prime}+r^{\prime})^2][1-(p^{\prime}-r^{\prime})^2].
\label{P96}
\end{eqnarray}
With these transformations the line element takes the conformal to flat form
\begin{equation}
ds^2=\frac{4L^2a^2(t)}{[1-(p^{\prime}+r^{\prime})^2][1-(p^{\prime}-r^{\prime})^2]}\left[-dp^{\prime 2}+dr^{\prime 2} +r^{\prime 2}d\theta^2+r^{\prime 2} \sin^2\theta d\phi^2\right].
\label{P97}
\end{equation}
As with the $K>0$ case, it is more convenient to continue to work in polar coordinates, and so we turn now to a fluctuation analysis in the polar coordinate system.

\subsection{Polar Coordinate Considerations}

To provide a comprehensive discussion of both $K>0$ and $K<0$ cosmologies we introduce a generic metric of the form
\begin{equation}
ds^2=\Omega^2(p,r)\left[-dp^{2}+dr^{ 2} +r^{ 2}d\theta^2+r^{2} \sin^2\theta d\phi^2\right],
\label{P98}
\end{equation}
with an a general conformal factor that depends on $p$ and $r$. For this metric we determine Christoffel symbols of the form
\begin{eqnarray}
&&\Gamma^{p}_{pp}=\Omega^{-1}\dot{\Omega},
\qquad \Gamma^{p}_{rr}=\Omega^{-1}\dot{\Omega},
\qquad \Gamma^{p}_{\theta\theta}=r^2\Omega^{-1}\dot{\Omega}, 
\qquad \Gamma^{p}_{\phi\phi}=r^2\sin^2\theta~\Omega^{-1}\dot{\Omega},
\nonumber\\
&&\Gamma^{r}_{pp}=\Omega^{-1}\Omega^{\prime},
\qquad \Gamma^{r}_{rr}=\Omega^{-1}\Omega^{\prime},
\qquad \Gamma^{r}_{\theta\theta}=-r-r^2\Omega^{-1}\Omega^{\prime}, 
\qquad \Gamma^{r}_{\phi\phi}=-\sin^2\theta(r+r^2\Omega^{-1}\Omega^{\prime}),
\nonumber\\
&&\Gamma^{p}_{pr}=\Omega^{-1}\Omega^{\prime},
\qquad \Gamma^{r}_{pr}=\Omega^{-1}\dot{\Omega},
\qquad \Gamma^{\theta}_{p\theta}=\Omega^{-1}\dot{\Omega}, 
\qquad \Gamma^{\theta}_{r\theta}=r^{-1}+\Omega^{-1}\Omega^{\prime},
\nonumber\\
&&\Gamma^{\phi}_{p\phi}=\Omega^{-1}\dot{\Omega},
\qquad \Gamma^{\phi}_{r\phi}=r^{-1}+\Omega^{-1}\Omega^{\prime},\qquad
\Gamma^{\theta}_{\phi\phi}=-\sin\theta \cos\theta,
\qquad \Gamma^{\phi}_{\theta\phi}=\cot \theta,
\label{P99}
\end{eqnarray}
where $\dot{\Omega}=\partial_{p}\Omega$ and $\Omega^{\prime}=\partial_{r}\Omega$.

Given the $\Omega^{-1}\bar{K}^{\mu\nu}\partial_{\nu}\Omega$ term in (\ref{P24}) and the fact that $\Omega$ is now to depend on $p$ and $r$, we see that we will be able to maintain the transverse gauge under a conformal transformation in a background geometry with such an $\Omega$ if we take all $\bar{K}_{p\mu}$ and $\bar{K}_{r\mu}$ components to vanish. With $\bar{K}_{\mu\nu}$ being traceless this will leave us with just two independent components, $\bar{K}_{\theta\phi}$ and $\bar{K}_{\theta\theta}=-\bar{K}_{\phi\phi}/\sin^2\theta$, just as needed for tensor fluctuations. (Transverse-traceless conditions in the sector of $\bar{K}_{\mu\nu}$ with angular indices  are the polar coordinate analogs of the Cartesian $\eta^{ij}\partial_{j}\bar{K}_{ik}=0$, $\eta^{ij}\bar{K}_{ij}=0$.) With the Christoffel symbols given above, one finds that when one takes all $\bar{K}_{p\mu}$ and $\bar{K}_{r\mu}$ components to vanish, the four components of $\overline{\nabla^{\nu}K_{\mu\nu}}$ evaluate to
\begin{eqnarray}
&&\overline{\nabla^{\nu}K_{p\nu}}=-\frac{\dot{\Omega}}{\Omega^3r^2}\left[\bar{K}_{\theta\theta}
+\frac{\bar{K}_{\phi\phi}}{\sin^2\theta}\right],
\nonumber\\
&&\overline{\nabla^{\nu}K_{r\nu}}=-\frac{1}{\Omega^2r^2}\left(\frac{1}{r}+\frac{\Omega^{\prime}}{\Omega}\right)
\left[\bar{K}_{\theta\theta}
+\frac{\bar{K}_{\phi\phi}}{\sin^2\theta}\right],
\nonumber\\
&&\overline{\nabla^{\nu}K_{\theta\nu}}=
\frac{1}{\Omega^2r^2}\left[\partial_{\theta}\bar{K}_{\theta\theta}
+\frac{1}{\sin^2\theta}\partial_{\phi}\bar{K}_{\theta\phi}
+\frac{\cos\theta}{\sin\theta}\bar{K}_{\theta\theta}
-\frac{\cos\theta}{\sin^3\theta}\bar{K}_{\phi\phi}\right],
\nonumber\\
&&\overline{\nabla^{\nu}K_{\phi\nu}}=
\frac{1}{\Omega^2r^2}\left[\partial_{\theta}\bar{K}_{\theta\phi}
+\frac{1}{\sin^2\theta}\partial_{\phi}\bar{K}_{\phi\phi}
+\frac{\cos\theta}{\sin\theta}\bar{K}_{\theta\phi}\right].
\label{P100}
\end{eqnarray}
Thus with the trace condition, we see that we can maintain $\overline{\nabla^{\nu}K_{\mu\nu}}=0$ for all four values of $\mu$ if we require that $\bar{K}_{\theta\theta}$, $ \bar{K}_{\theta\phi}$ and $ \bar{K}_{\phi\phi}$ be independent of $\phi$, and that their  $\theta$ dependences be given as
\begin{equation}
\bar{K}_{\theta\theta} \sim \sin^{-2}\theta,\qquad \bar{K}_{\theta\phi} \sim \sin^{-1}\theta,\qquad  \bar{K}_{\phi\phi} \sim \sin^{0}\theta.
\label{P101}
\end{equation}

In a  polar coordinate flat metric of the form 
\begin{equation}
ds^2=-dp^{2}+dr^{ 2} +r^{ 2}d\theta^2+r^{2} \sin^2\theta d\phi^2,
\label{P102}
\end{equation}
the non-vanishing Christoffel symbols are given by 
\begin{eqnarray}
&&\Gamma^{r}_{\theta\theta}=-r, 
\qquad \Gamma^{r}_{\phi\phi}=-r\sin^2\theta,
\qquad \Gamma^{\theta}_{r\theta}=r^{-1},
\nonumber\\
&&\Gamma^{\phi}_{r\phi}=r^{-1},\qquad
\Gamma^{\theta}_{\phi\phi}=-\sin\theta~\cos\theta,
\qquad \Gamma^{\phi}_{\theta\phi}=\cot \theta.
\label{P103}
\end{eqnarray}
Modes that obey $\nabla^{\nu}K_{\mu\nu}=0$, $g_{(0)}^{\mu\nu}K_{\mu\nu}=0$, $K_{p\mu}=0$, $K_{r\mu}=0$  are then found to obey 
\begin{eqnarray}
&&\nabla^{\nu}K_{p\nu}=0,
\nonumber\\
&&\nabla^{\nu}K_{r\nu}=-\frac{1}{r^3}\left[K_{\theta\theta}
+\frac{K_{\phi\phi}}{\sin^2\theta}\right]=0,
\nonumber\\
&&\nabla^{\nu}K_{\theta\nu}=
\frac{1}{r^2}\left[\partial_{\theta}K_{\theta\theta}
+\frac{1}{\sin^2\theta}\partial_{\phi}K_{\theta\phi}
+\frac{\cos\theta}{\sin\theta}K_{\theta\theta}
-\frac{\cos\theta}{\sin^3\theta}K_{\phi\phi}\right]=0,
\nonumber\\
&&\nabla^{\nu}K_{\phi\nu}=
\frac{1}{r^2}\left[\partial_{\theta}K_{\theta\phi}
+\frac{1}{\sin^2\theta}\partial_{\phi}K_{\phi\phi}
+\frac{\cos\theta}{\sin\theta}K_{\theta\phi}\right]=0,
\label{P104}
\end{eqnarray}
and thus obey 
\begin{equation}
K_{\theta\theta} \sim \sin^{-2}\theta,\qquad K_{\theta\phi} \sim \sin^{-1}\theta,\qquad
K_{\phi\phi} \sim \sin^{0}\theta.
\label{P105}
\end{equation}
Comparing (\ref{P104}) and (\ref{P105}) with (\ref{P100}) and (\ref{P101}), we see that under the conformal transformation with $\Omega(p,r)$ the transverse gauge condition is indeed preserved, just as required. With (\ref{P101}) we have obtained all the constraints on the fluctuations that can be derived purely via kinematic considerations, and to determine their dependence on $p$ and $r$ we proceed now to the fluctuation equations that $\bar{K}_{\theta\theta}$ and $ \bar{K}_{\theta\phi}$ obey.

\subsection{Fluctuations in the Polar Coordinate System}

To determine $\bar{K}_{\theta\theta}$ and $ \bar{K}_{\theta\phi}$  we need to determine $K_{\theta\theta}$ and $K_{\theta\phi}$ and then make a conformal transformation. The fluctuations $K_{\theta\theta}$ and $K_{\theta\phi}$ are associated with fluctuations around the metric with the line element given in (\ref{P102}). We thus need to determine $\delta W_{\mu\nu}$ in the (\ref{P102}) background. While this background is flat it is not in a Cartesian coordinate system, and hence we cannot use (\ref{P50}). However, our derivation of (\ref{P49}) given above only used generic properties of variations such as $\delta [\nabla_{\sigma}\nabla_{\lambda}R_{\mu\nu}]$ and actually made no specific reference to the metric other than in the use of (\ref{P26}) for the de Sitter form for the Riemann and Ricci tensors. Thus if we simply set $H=0$ in (\ref{P49}) we will obtain an expression for  $\delta W_{\mu\nu}$ that will hold for a flat geometry as written in an arbitrary curvilinear coordinate system, viz.
\begin{eqnarray}
\delta W_{\mu\nu}&=&\frac{1}{2}\nabla_{\alpha}\nabla^{\alpha}\nabla_{\beta}\nabla^{\beta}K_{\mu\nu}
-\frac{1}{2}\nabla_{\beta}\nabla^{\beta}[
\nabla_{\mu}\nabla_{\lambda}K^{\lambda}_{\phantom{\lambda}\nu}
+\nabla_{\nu}\nabla_{\lambda}K^{\lambda}_{\phantom{\lambda}\mu}]
\nonumber\\
&+&\frac{1}{6}[g_{\mu\nu}\nabla_{\alpha}\nabla^{\alpha}+2\nabla_{\mu}\nabla_{\nu}
]\nabla_{\kappa}\nabla_{\lambda}K^{\kappa\lambda}.
\label{P106}
\end{eqnarray}
If we now restrict to the transverse gauge, $\delta W_{\mu\nu}$ reduces to 
\begin{equation}
\delta W_{\mu\nu}=\frac{1}{2}\nabla_{\alpha}\nabla^{\alpha}\nabla_{\beta}\nabla^{\beta}K_{\mu\nu},
\label{107}
\end{equation}
and thus all we need to do is evaluate $\nabla_{\beta}\nabla^{\beta}A_{\mu\nu}$ for $A_{\mu\nu}=K_{\mu\nu}$ and $A_{\mu\nu}=\nabla_{\beta}\nabla^{\beta}K_{\mu\nu}$ in the (\ref{P102}) background.

With the general $\nabla_{\beta}\nabla^{\beta}A_{\mu\nu}$ being given in (\ref{P52}), we only need to evaluate (\ref{P52}) using the Christoffel symbols given in (\ref{P103}). On setting $A_{p\mu}=0$ and $A_{r\mu}=0$ in the right-hand side of (\ref{P52}) but without imposing any other other conditions on $A_{\mu\nu}$, following some algebra we obtain
\begin{eqnarray}
&&\nabla_{\beta}\nabla^{\beta}A_{0\mu}=0,
\nonumber\\
&&\nabla_{\beta}\nabla^{\beta}A_{rr}=\frac{2}{r^4}\left[A_{\theta\theta}
+\frac{A_{\phi\phi}}{\sin^2\theta}\right],
\nonumber\\
&&\nabla_{\beta}\nabla^{\beta}A_{r\theta}=\frac{2}{r^3}\left[\frac{\cos\theta}{\sin^3\theta}A_{\phi\phi}-\frac{\cos\theta}{\sin\theta}A_{\theta\theta}-\partial_{\theta}A_{\theta\theta}
-\frac{1}{\sin^2\theta}\partial_{\phi}A_{\phi\theta}\right],
\nonumber\\
&&\nabla_{\beta}\nabla^{\beta}A_{r\phi}=-\frac{2}{r^3}\left[\frac{1}{\sin^2\theta}\partial_{\phi}A_{\phi\phi}
+\frac{\cos\theta}{\sin\theta}A_{\theta\phi}+\partial_{\theta}A_{\theta\phi}
\right].
\label{P108}
\end{eqnarray}
Then with $A_{\theta\theta}+A_{\phi\phi}/\sin^2\theta=0$, $A_{\theta\theta}\sim \sin^{-2}\theta$, $A_{\theta\phi}\sim \sin^{-1}\theta$, $A_{\phi\phi}\sim \sin^{0}\theta$, $\partial_{\phi}A_{\theta\theta}=0$, $\partial_{\phi}A_{\phi\theta}=0$, $\partial_{\phi}A_{\phi\phi}=0$, we find that we can consistently maintain $\nabla_{\beta}\nabla^{\beta}A_{p\mu}=0$, $\nabla_{\beta}\nabla^{\beta}A_{r\mu}=0$, just as required. Finally, for $A_{\theta\theta}$, $A_{\theta\phi}$, and $A_{\theta\phi}$, we obtain
\begin{eqnarray}
&&\nabla_{\beta}\nabla^{\beta}A_{\theta\theta}=\left[D -\frac{4}{r}\frac{\partial}{\partial r}
-\frac{4\cos^2\theta}{r^2\sin^2\theta}\right] A_{\theta\theta}
-\frac{2\cos^2\theta}{r^2\sin^3\theta}\partial_{\phi} A_{\phi\theta},
\nonumber\\
&&\nabla_{\beta}\nabla^{\beta}A_{\theta\phi}=\left[D -\frac{4}{r}\frac{\partial}{\partial r}
+\frac{1}{r^2}
-\frac{3\cos^2\theta}{r^2\sin^2\theta}
-\frac{2\cos\theta}{r^2\sin\theta}\frac{\partial}{\partial \theta}\right] A_{\theta\phi}
+\frac{2\cos\theta}{r^2\sin\theta}\partial_{\phi}\left[ A_{\theta\theta}
-\frac{A_{\phi\phi}}{\sin^2\theta}\right],
\nonumber\\
&&\nabla_{\beta}\nabla^{\beta}A_{\phi\phi}=\left[D -\frac{4}{r}\frac{\partial}{\partial r}
+\frac{2}{r^2\sin^2\theta}
-\frac{4\cos\theta}{r^2\sin\theta}\frac{\partial}{\partial \theta}\right] A_{\phi\phi}
+\frac{2\cos^2\theta}{r^2}A_{\theta\theta}
+\frac{4\cos\theta}{r^2\sin\theta}\partial_{\phi} A_{\theta\phi},
\label{P109}
\end{eqnarray}
where
\begin{equation}
D=g^{RL}[\partial_{L}\partial_{R}-\Gamma^{S}_{LR}\partial_{S}]
=-\frac{\partial^2}{\partial p^2}
+\frac{\partial^2}{\partial r^2} +\frac{2}{r}\frac{\partial}{\partial r}
+\frac{1}{r^2}\frac{\partial^2}{\partial \theta^2}
+\frac{\cos\theta}{r^2\sin\theta}\frac{\partial}{\partial \theta}
+\frac{1}{r^2\sin^2\theta}\frac{\partial^2}{\partial \phi^2}.
\label{P110}
\end{equation}
When we impose  $\partial_{\phi}A_{\theta\theta}=0$, $\partial_{\phi}A_{\phi\theta}=0$, $\partial_{\phi}A_{\phi\phi}=0$, $A_{\theta\theta}+A_{\phi\phi}/\sin^2\theta=0$, we find that (\ref{P109}) becomes diagonal in its indices. Finally when we set $A_{\theta\theta}\sim \sin^{-2}\theta$, $A_{\theta\phi}\sim \sin^{-1}\theta$, $A_{\phi\phi}\sim \sin^{0}\theta$, we find that the dependence on $\theta$ drops out, with (\ref{P109}) reducing to 
\begin{eqnarray}
&&\nabla_{\beta}\nabla^{\beta}A_{\theta\theta}=\left[-\frac{\partial^2}{\partial p^2}+\frac{\partial^2}{\partial r^2} -\frac{2}{r}\frac{\partial}{\partial r}+\frac{2}{r^2}\right]A_{\theta\theta},
\nonumber\\
&&\nabla_{\beta}\nabla^{\beta}A_{\theta\phi}=\left[-\frac{\partial^2}{\partial p^2}+\frac{\partial^2}{\partial r^2} -\frac{2}{r}\frac{\partial}{\partial r}+\frac{2}{r^2}\right]A_{\theta\phi},
\nonumber\\
&&\nabla_{\beta}\nabla^{\beta}A_{\phi\phi}=\left[-\frac{\partial^2}{\partial p^2}+\frac{\partial^2}{\partial r^2} -\frac{2}{r}\frac{\partial}{\partial r}+\frac{2}{r^2}\right]A_{\phi\phi}.
\label{P111}
\end{eqnarray}
This then is our key result in the polar coordinate case.

In parallel to the discussion of the derivation of (\ref{P74}) given above, we note that 
\begin{equation}
\left[-\frac{\partial^2}{\partial p^2}+\frac{\partial^2}{\partial r^2} -\frac{2}{r}\frac{\partial}{\partial r}+\frac{2}{r^2}\right][(a+bp)re^{iqr-iqp}]=2iqbre^{iqr-iqp},
\label{P112}
\end{equation}
and that 
\begin{equation}
\left[-\frac{\partial^2}{\partial p^2}+\frac{\partial^2}{\partial r^2} -\frac{2}{r}\frac{\partial}{\partial r}+\frac{2}{r^2}\right]re^{iqr-iqp}=0.
\label{P113}
\end{equation}
Thus we find exact positive frequency solutions to $\nabla_{\alpha}\nabla^{\alpha}\nabla_{\beta}\nabla^{\beta}A_{\mu\nu}=0$ of the form
\begin{eqnarray}
&&A_{\theta\theta}=[\alpha_{\theta\theta}+\beta_{\theta\theta}p]re^{iqr-iqp}\sin^{-2}\theta,
\nonumber\\
&&A_{\theta\phi}=[\alpha_{\theta\phi}+\beta_{\theta\phi}p]re^{iqr-iqp}\sin^{-1}\theta,
\nonumber\\
&&A_{\phi\phi}=-[\alpha_{\theta\theta}+\beta_{\theta\theta}p]re^{iqr-iqp}.
\label{P114}
\end{eqnarray}
as labeled in terms of a radial momentum $q$. 

With (\ref{P114}) holding for $A_{\mu\nu}=K_{\mu\nu}$, we can now construct $\bar{K}_{\mu\nu}$ by conformal transformation. With reference to the line element given in (\ref{P87}) and (\ref{P90}), we find that since coordinate transformations in the temporal and radial coordinate sectors do not affect the angular components of either $\bar{K}_{\mu\nu}$ or $\bar{g}^{(0)}_{\mu\nu}$, for a topologically closed $K>0$ RW geometry the positive frequency sector of $\bar{K}_{\theta\theta}$ is given by 
\begin{eqnarray}
\bar{K}_{\theta\theta}(t,r,\theta,\phi)&=&L^2a^2(t)\cos^2[(p+\chi)/2]\cos^2[(p-\chi)/2]
\nonumber\\
&\times&\bigg{[}2\alpha^{(+)}_{\theta\theta}+\beta^{(+)}_{\theta\theta}\tan[(p+\chi)/2]+\beta^{(+)}_{\theta\theta}\tan[(p-\chi)/2]\bigg{]}
\nonumber\\
&\times&\bigg{[}\tan[(p+\chi)/2]-\tan[(p-\chi)/2]\bigg{]}\frac{\exp[-iq\tan[(p-\chi)/2]]}{\sin^{2}\theta},
\label{P115}
\end{eqnarray}
where
\begin{equation}
p=\frac{\tau}{L}=\frac{1}{L}\int \frac{dt}{a(t)},\qquad \chi={\rm arc~sin}\left(\frac{r}{L}\right),
\label{P116}
\end{equation}
with analogous expressions holding for $\bar{K}_{\theta\phi}(t,r,\theta,\phi)$ and $\bar{K}_{\phi\phi}(t,r,\theta,\phi)$. Similarly, with reference to the line element given in (\ref{P87}) and (\ref{P94}), we find that for a topologically open $K<0$ RW geometry the positive frequency sector of $\bar{K}_{\theta\theta}$ is given by 
\begin{eqnarray}
\bar{K}_{\theta\theta}(t,r,\theta,\phi)&=&L^2a^2(t){\rm cosh}^2[(p+\chi)/2]{\rm cosh}^2[(p-\chi)/2]
\nonumber \\
&\times&\bigg{[}2\alpha^{(-)}_{\theta\theta}+\beta^{(-)}_{\theta\theta}{\rm tanh}[(p+\chi)/2]+\beta^{(-)}_{\theta\theta}{\rm tanh}[(p-\chi)/2]\bigg{]}
\nonumber\\
&\times&\bigg{[}{\rm tanh}[(p+\chi)/2]-{\rm tanh}[(p-\chi)/2]\bigg{]}\frac{\exp[-iq{\rm \tanh}[(p-\chi)/2]]}{\sin^{2}\theta},
\label{P117}
\end{eqnarray}
where
\begin{equation}
p=\frac{\tau}{L}=\frac{1}{L}\int \frac{dt}{a(t)},\qquad \chi={\rm arc~sinh}\left(\frac{r}{L}\right),
\label{P118}
\end{equation}
with analogous expressions holding for $\bar{K}_{\theta\phi}(t,r,\theta,\phi)$ and $\bar{K}_{\phi\phi}(t,r,\theta,\phi)$.  (The various $\alpha^{(\pm)}_{\mu\nu}$ and $\beta^{(\pm)}_{\mu\nu}$ coefficients that appear in (\ref{P115}) and (\ref{P117}) are constants.) With (\ref{P115}) and (\ref{P117}) we have thus obtained exact, closed form expressions for the first-order tensor fluctuations around the spatially non-flat $K>0$ and $K<0$ RW geometries in the conformal gravity theory.

\subsection{Explicit Forms for $a(t)$}

To explore the implications of the structure that we have found for the fluctuations, we need to provide an explicit form for the background $a(t)$ in each of the various cases. To this end we recall the conformal cosmology studies described in 
\cite{Mannheim1991a,Mannheim1998,Mannheim2001,Mannheim2006}. In those studies the matter field energy-momentum tensor that is needed to fix $a(t)$ was taken to be composed of a perfect fluid of fermions together with a conformally coupled scalar field  order parameter  that generates the needed spontaneous breakdown of the conformal symmetry. The scalar field acts as the  expectation value of a fermion bilinear condensate in a spontaneously broken vacuum and dynamically generates particle masses.

To illustrate the central role that is played by the trace of the matter field energy-momentum tensor $T^{\rm M}_{\mu \nu}$  in these conformal studies, and especially to see what happens to the trace when particle masses are generated spontaneously by a scalar field $S(x)$ (so that macroscopic inhomogeneities built out of such massive particles can then form), it is convenient to focus first on each of the individual fermions in the fluid. For each such individual fermion $\psi(x)$ the most general conformal invariant and coordinate invariant matter field sector action is of the form 
\begin{equation}
I_{\rm M}=-\int d^4x(-g)^{1/2}\left[
i\bar{\psi}\gamma^{\mu}(x)[\partial_\mu+\Gamma_\mu(x)]             
\psi -hS\bar{\psi}\psi
+\frac{1}{2}S^{;\mu}
S_{;\mu}-\frac{1}{12}S^2R^\mu_{\phantom{\mu}\mu}+\lambda S^4 \right],
\label{P119}
\end{equation}                                 
where $\Gamma_\mu(x)$ is the fermion spin connection and the coefficients $h$ and $\lambda$ are dimensionless. In (\ref{P119}) there is no intrinsic or mechanical $m\bar{\psi}\psi$ mass term since such a mass term would violate the conformal symmetry. On using the matter field sector equations of motion 
\begin{eqnarray}
i\gamma^{\mu}(x)[\partial_{\mu} +\Gamma_\mu(x)]                              
\psi - h S \psi &=& 0,
\nonumber \\
S^{;\mu}_{\phantom{\mu};\mu}+\frac{1}{6}SR^\mu_{\phantom{\mu}\mu}
-4\lambda S^3 +h\bar{\psi}\psi &=& 0,
\label{P120}
\end{eqnarray}
one finds that in a configuration in which the scalar field takes a constant value $S_0$, the fermion acquires a mass $m=hS_0$. In such a configuration the entire matter field energy-momentum tensor takes the form \cite{Mannheim2006} 
\begin{equation}
T^{\rm M}_{\mu \nu} = T^{{\rm kin}}_{\mu \nu} 
-\frac{1}{6}S_0^2\left(R_{\mu\nu}
-\frac{1}{2}g_{\mu\nu}R^\alpha_{\phantom{\alpha}\alpha}\right)         
-g_{\mu\nu}\lambda S_0^4,
\label{P121}
\end{equation}                                 
where $T^{{\rm kin}}_{\mu \nu} =i \bar{\psi} \gamma_{\mu}(x)[\partial_{\nu}+\Gamma_\nu(x)]  \psi$ is the kinetic energy of the fermion. As we see, even though $T^{{\rm kin}}_{\mu \nu}$ itself is not traceless in the presence of dynamical mass generation ($g^{\mu\nu}T^{{\rm kin}}_{\mu \nu} =hS_0 \bar{\psi}\psi$), because of the scalar field equation of motion the full  $T^{\rm M}_{\mu \nu}$ is traceless ($g^{\mu\nu}T^{\rm M}_{\mu \nu}=0$). Thus while mechanical masses would violate the conformal symmetry and lead to the non-vanishing of the trace of the matter field energy-momentum tensor, in the dynamical case the scalar field that generates masses for the fermions also carries energy and momentum itself, and its contribution to $T^{\rm M}_{\mu \nu}$ then serves  \cite{Mannheim2006} to maintain $g^{\mu\nu}T^{\rm M}_{\mu \nu}=0$. Since $W^{\mu\nu}$ is traceless, we see that even in the presence of dynamical mass generation for particles (and thus of the inhomogeneous fluctuations that are built out of them) $T_{\rm M}^{\mu \nu}$ can serve as the source for $W^{\mu\nu}$, just as exhibited in (\ref{P3}). As we see, conformal invariance is not just relevant at energies well above particle masses, viz. at energies in which the vacuum expectation value of the scalar field can be taken to be zero, even at  low energies where one is in the spontaneously broken phase the conformal structure has residual effects, with $g^{\mu\nu}T^{\rm M}_{\mu \nu}$ still vanishing. 

While only the trace of $T^{\rm M}_{\mu \nu}$ would vanish in the presence of inhomogeneous fluctuations (viz. inhomogeneous geometries in which both $W^{\mu\nu}$ and $C^{\lambda\mu\nu\kappa}$ are non-zero), in the background cosmology itself both $C^{\lambda\mu\nu\kappa}$ and all 10 components of the gravitational $W^{\mu\nu}$ would vanish identically since the background geometry is conformal to flat. Thus in conformal cosmology where $T^{{\rm kin}}_{\mu \nu}$ has the form of an entire perfect fluid of the individual fermions, viz. $T^{{\rm kin}}_{\mu \nu}=(\rho_m+p_m)U_{\mu}U_{\nu}+p_mg_{\mu\nu}$, each of the 10 components of the matter field energy-momentum tensor must obey
\begin{equation}
T^{\rm M}_{\mu \nu}=0,
\label{P122}
\end{equation}                                 
with the Universe thus being created out of nothing.  

Despite the fact that $T^{\rm M}_{\mu \nu}$ is zero in conformal cosmology, because of the induced Einstein tensor and cosmological constant terms present in (\ref{P121}) its vanishing can be achieved non-trivially, with there being five possible ways to do so. (While $\lambda<0$ is to be favored since it corresponds to a lowering of the free energy of the scalar field order parameter \cite{Mannheim2006}, we consider all $\lambda$ values here.) Specifically, in terms of the illustrative  perfect fluid with $\rho_m=3p_m=A/a^4(t)$ where $A>0$, the solutions are  found to depend on the signs of $K$ and $\alpha$, where
\begin{equation}
\alpha=-2\lambda S_0^2.
\label{P123}
\end{equation}                                 
In terms of the convenient parameter
\begin{equation}
\beta=\left(1- \frac{16A\lambda}{K^2}\right)^{1/2},
\label{P124}
\end{equation}                                 
the five separate solutions are given as 
\cite{Mannheim1991a,Mannheim1998,Mannheim2001,Mannheim2006}
\begin{eqnarray}
a^2(t,\alpha<0,K<0)&=&\frac{K(1-\beta)}{2\alpha}+
\frac{K\beta \sin^2 ((-\alpha)^{1/2} t)}{\alpha},
\nonumber \\
a^2(t,\alpha=0,K<0)&=&-\frac{2A}{K  S_0^2}-Kt^2,
\nonumber \\
a^2(t,\alpha>0,K<0)&=&-\frac{K(\beta-1)}{2\alpha}
-\frac{K\beta {\rm sinh}^2 (\alpha^{1/2} t)}{\alpha},
\nonumber \\
a^2(t,\alpha > 0,K=0)&=&\left(-\frac{A}{\lambda S_0^4}\right)^{1/2}
{\rm cosh}(2\alpha^{1/2}t),
\nonumber \\
a^2(t,\alpha > 0,K>0)&=&\frac{K(1+\beta)}{2\alpha}+
\frac{K\beta {\rm sinh}^2 (\alpha^{1/2} t)}{\alpha}.
\label{P125}
\end{eqnarray}

While one could in principle insert of all these forms for $a(t)$ into the various expressions for the fluctuations that we found above, the ones of most interest to conformal gravity are those with $K<0$. Specifically, on theoretical grounds one can show \cite{Mannheim2000a} that above the critical temperature at which $S_0$ becomes non-zero, the only non-trivial solutions to the resulting $T^{{\rm kin}}_{\mu \nu} =0$ are those with $K<0$, with the positive energy density of the matter field being cancelled by the negative energy density of a $K<0$ gravitational field to which it is coupled, and with the global topology of the Universe being fixed once and for all before it undergoes any phase transitions as it cools. Secondly, as we discuss below, galactic rotation curves probe the structure of the global cosmology (these global effects replace the need for local dark matter within galaxies) and lead \cite{Mannheim1989,Mannheim1997} to negative $K$. Thirdly, in the fits to the accelerating Universe data, again it is $K<0$ that is favored \cite{Mannheim2003,Mannheim2006}.

While the supernovae fits favor both $K<0$ and $\alpha>0$ (viz. the theoretically favored $\lambda<0$), there is not a great deal of statistical difference between those fits and ones with $K<0$ and $\alpha=0$. Since the associated $\alpha=0$ $a(t)$ is more tractable, for our purposes here we shall restrict to the $K<0$, $\alpha=0$ solution. On recalling that $K=-1/L^2$, in this solution we obtain 
\begin{equation}
p=\frac{\tau}{L}=(-K)^{1/2}\int \frac{dt}{a(t,\alpha=0,K<0)}={\rm arc sinh}\left(\frac{-KS_0t}{(2A)^{1/2}}\right).
\label{P126}
\end{equation}
Inspection of (\ref{P117}) shows that for lightlike configurations with $p=\chi$ (i.e. along a null geodesic of the geometry) tensor fluctuations such as $\bar{K}_{\theta\theta}(t,r,\theta,\phi)$ grow as $t^4$, while in those timelike configurations for which $p \gg \chi$ (i.e. for $-KS_0t/(2A)^{1/2} \gg r$) the fluctuations grow as $t^6$. This is not only quite rapid growth, it is altogether more rapid than the growth obtained for tensor fluctuations around an RW geometry in the standard second-order theory.

\section{General Comments}
\label{S6}
\subsection{Implications for Cosmology}

In this paper we have provided the initial groundwork needed to develop conformal cosmological perturbation theory. Much of course remains to be done, especially in determining the behavior of the matter field fluctuations and in discussing fluctuations in the gravitational sector that are not of the spatially transverse and traceless tensor type discussed here. Such a more detailed analysis is needed in order to be able to address the anisotropy of the cosmic microwave background and the growth of large scale structure in the conformal theory. One of the potential benefits of such a study could be in the behavior of the cosmic microwave background at large angles where the inflationary Universe \cite{Guth1981} gives too much power (see e.g. \cite{Copi2010} and references therein for a recent discussion of the current status of the problem).  

To be more specific, we note that if we consider a generic expansion rate of the form $a(t)\sim t^n$ starting off at some initial time $t_i$, the horizon size integral $d_{\rm H}(t)=\int_{t_i}^t dt/a(t)$ will behave as $[t^{1-n}/(1-n)]|_{t_i}^t$. If the expansion is an adiabatic one with $a(t)=1/T(t)$, then $d_{\rm H}(t)=[T_{\rm MAX}^{(n-1)/n}-T(t)^{(n-1)/n}]/(n-1)$ where $T_{\rm MAX}=T(t_i)$. Consequently, if the recombination temperature is altogether smaller than the initial temperature $T_{\rm MAX}$, the horizon size at recombination will become very large if $n \geq 1$, but will be very small if $0<n<1$. Background RW cosmologies with $n \geq 1$ will thus have no horizon problem, while those with $0<n<1$ will \cite{footnote1}. Since the standard cosmology leads to a $t^{1/2}$ behavior in the radiation era, the standard early Universe  RW cosmology has a horizon problem. With inflation leading to an exponential $a(t)\sim e^{Ht}$ expansion rate, such an $a(t)$ grows faster than any power and immediately solves the horizon problem. However, inflation solves the horizon problem by a lot more than is needed (just $a(t)\sim t^n$ with $n \geq 1$ is needed) and could thus be getting much further outside the standard cosmology horizon than is necessary, and potentially be producing too much power on large angles. In contrast to the standard RW cosmology, the expansion rate of conformal RW cosmology is such that it does not lead to any horizon problem \cite{footnote2} (though it does so at the expense of a possible nucleosynthesis problem \cite{Knox1993,Elizondo1994,Lohiya1999,Sethi1999}). And as we saw above, the conformal cosmology fluctuations grow much faster than the ones obtained in a standard RW cosmology but less rapidly than the ones obtained in an inflationary cosmology. It would thus be of interest to see if conformal cosmology strikes the right balance of having a background that expands fast enough to have no horizon problem, while having fluctuations that grow fast enough to produce a detectable anisotropy at the time of recombination but not so fast as to produce too much power at large angles. 

While a more detailed discussion of the anisotropy of the cosmic microwave background in the conformal case must await an explicit calculation, we should point out here that the need for a negative $K$ in the conformal case is not necessarily in conflict with the flatness found in data analyses based on the standard inflationary model. Specifically, what is determined in current cosmic microwave background analyses is the size of fluctuations as measured at recombination compared to the ``standard ruler'' size to which they are expected to grow to after starting in the inflationary era. Since the growth rate of early Universe fluctuations is quite different in the conformal case, the standard ruler size to which they would grow by recombination could be radically different from the size that occurs in the standard inflationary model, with a measurement of that size potentially leading to a totally different inferred value for $K$.

\subsection{Implications for Galactic Rotation Curves}

As we had noted above, galaxies are inhomogeneities, with the recent work of 
\cite{Mannheim2010b,Mannheim2010c,O'Brien2011} on galactic rotation curves actually having a connection to the fluctuation issue.  In the original study of the geometry associated with static, spherically symmetric sources in the conformal theory, it was noted \cite{Mannheim1989,Mannheim1994} that via a sequence of conformal and coordinate transformations the line element in a static, spherically symmetric geometry could be brought to the form
\begin{equation}
ds^2=-B(r)c^2dt^2+\frac{dr^2}{B(r)}+r^2d\theta^2+r^2\sin^2\theta~d\phi^2,
\label{P127}
\end{equation}
to thus be expressed in terms of a single metric coefficient $B(r)$. In terms of this line element the exact fourth-order equation of motion given in (\ref{P3}) reduces to the remarkably simple fourth-order Poisson form
\begin{equation}
\frac{3}{B(r)}(W^{0}_{\phantom{0}0}-W^{r}_{\phantom{r}r})=\nabla^4B=\frac{d^4B}{dr^4}+\frac{4}{r}\frac{d^3B}{dr^3}=\frac{3}{4\alpha_gB(r)}(T^{0}_{\phantom{0}0}-T^{r}_{\phantom{r}r})=f(r).
\label{P128}
\end{equation}
Solutions to (\ref{P128}) are of the form
\begin{eqnarray}
B(r)&=& -\frac{r}{2}\int_0^r
dr^{\prime}\,r^{\prime 2}f(r^{\prime})
-\frac{1}{6r}\int_0^r
dr^{\prime}\,r^{\prime 4}f(r^{\prime})
\nonumber\\
&&-\frac{1}{2}\int_r^{\infty}
dr^{\prime}\,r^{\prime 3}f(r^{\prime})
-\frac{r^2}{6}\int_r^{\infty}
dr^{\prime}\,r^{\prime }f(r^{\prime}) +\hat{B}(r),
\label{P129}
\end{eqnarray}                                 
where $\hat{B}(r)$ obeys $\nabla^4\hat{B}(r)=0$. 

For a localized source function $f(r)$ that is restricted to the $0\leq r \leq r_0$ region, only the first two integrals in  (\ref{P129}) contribute to $B(r>r_0)$ and yield (on conveniently setting $\hat{B}(r)=1$) 
\begin{equation}
B(r>r_0)=1-\frac{2\beta}{r}+\gamma r,
\label{P130}
\end{equation}
where
\begin{equation}
2\beta=\frac{1}{6}\int_0^{r_0}\,dr^{\prime}\,r^{\prime 4}f(r^{\prime}),\qquad \gamma= -\frac{1}{2}\int_0^{r_0}dr^{\prime}\,r^{\prime 2}f(r^{\prime}).
\label{P131}
\end{equation}
The term in (\ref{P130}) that grows linearly with $r$ is the spatial analog of the linear in time behavior found in (\ref{P74}), and leads to potentials that grow rather than fall at large distances. An integration of (\ref{P130}) over the material within the galaxy yields a net $B_{\rm LOC}(r)$ contribution to the galactic potential generated by the local material within the galaxy. 

However, unlike the situation in standard gravity, one cannot use the net local contribution $B_{\rm LOC}(r)$ alone (even with its linear term) to fit galactic rotation curves as one cannot ignore the linear potentials coming from inhomogeneous material outside of the galaxies. Rather the contribution coming from inhomogeneities  located outside the galaxy generates the last two integrals in (\ref{P129}). As discussed in \cite{Mannheim2010b,Mannheim2010c,O'Brien2011}, if the exterior $f(r)$ distribution starts at some typical cluster radius $r_{\rm clus}$, then in the $r_0\leq r \leq r_{\rm clus}$ region the contribution of the fourth integral acts like a quadratic de Sitter like potential with a strength 
\begin{equation}
\kappa =\frac{1}{6}\int_{r_{\rm clus}}^{\infty} dr^{\prime}\,r^{\prime }f(r^{\prime}).
\label{P132}
\end{equation}
(While the third integral in (\ref{P129}) does contribute to the potential, when the integration range starts at a fixed  $r_{\rm clus}$, this integral makes no contribution to the force and so we do not consider it any further.) Since the integral in (\ref{P132}) is independent of the galaxy of interest, its contribution is universal for all galaxies.

In addition to  (\ref{P132}) there is a further global contribution provided by material outside the galaxy, this one associated with non-trivial solutions to $\nabla^4\hat{B}(r)=0$, a thus necessarily homogeneous contribution in which $W_{\mu\nu}$ given in (\ref{P3}) is zero. As recognized in \cite{Mannheim1989,Mannheim1997}, this contribution is due to the background cosmology itself. Specifically,  we recall \cite{Mannheim1989,Mannheim1997} that a coordinate transformation of the form
\begin{equation}
\rho=\frac{4r}{2(1+\gamma_0r)^{1/2}+2 +\gamma_0 r},\qquad r= \frac{\rho}{(1-\gamma_0\rho/4)^2},\qquad \tau=\int dt\,a(t)
\label{P133}
\end{equation}                                 
effects the metric transformation
\begin{eqnarray}
&&-(1+\gamma_0r)c^2dt^2+\frac{dr^2}{(1+\gamma_0r)}+r^{ 2}d\theta^2+r^{2} \sin^2\theta d\phi^2
\nonumber \\
&&=\frac{1}{a^2(\tau)}\left(\frac{1+\gamma_0\rho/4}
{1-\gamma_0\rho/4}\right)^2
\left[-c^2d\tau^2+\frac{a^2(\tau)}{[1-\gamma_0^2\rho^2/16]^2}
\left(d\rho^2+\rho^2d\theta^2+\rho^2 \sin^2\theta d\phi^2\right)\right].
\label{P134}
\end{eqnarray} 
With the bracketed term on the right-hand side of  (\ref{P134}) being recognized as an RW geometry with an expressly negative definite $K=-\gamma_0^2/4$, we see that in the rest frame of a galaxy, a $K<0$ cosmological background acts as a universal linear potential $\gamma_0$ whose strength is independent of the galaxy of interest. Since as noted above conformal gravity precisely produces such a $K<0$ RW background, its effect on galactic rotation curves is to produce an effective universal linear potential term.

With the net local $B_{\rm LOC}(r)$ contribution is combined with the two universal global terms, the total weak gravity metric is then given by
\begin{equation}
B_{\rm TOT}(r)=B_{\rm LOC}(r)+\gamma_0 r -\kappa r^2.
\label{P135}
\end{equation}                                 
With  the luminous material in a spiral galaxy typically being distributed with a surface brightness of the form $\Sigma(R)=e^{-R/R_0}L/2\pi R_0^2$ where $L$ is the total luminosity, and with the potential produced by a single star being of the form $V^{*}(r)=[B^*(r)-1]c^2/2=-\beta^{*}c^2/r+\gamma^{*} c^2 r/2$ as per (\ref{P130}), the local galactic $B_{\rm LOC}(r)$ is readily computed \cite{Mannheim2006}, with rotation velocities then being given by the very compact formula
\begin{eqnarray}
v_{\rm TOT}^2(R)&=&
\frac{N^*\beta^*c^2 R^2}{2R_0^3}\left[I_0\left(\frac{R}{2R_0}
\right)K_0\left(\frac{R}{2R_0}\right)-
I_1\left(\frac{R}{2R_0}\right)
K_1\left(\frac{R}{2R_0}\right)\right]
\nonumber \\
&+&\frac{N^*\gamma^* c^2R^2}{2R_0}I_1\left(\frac{R}{2R_0}\right)
K_1\left(\frac{R}{2R_0}\right)+\frac{\gamma_0 c^2R}{2}-\kappa c^2R^2,
\label{P136}
\end{eqnarray} 
where $N^*M_{\odot}=M=(M/L)L$ is the total mass of the galaxy. Through the use of (\ref{P136}) and with no local material being included in $B_{\rm LOC}(r)$ other than the known luminous material within galaxies (i.e. no dark matter), successful fitting to no less than 138 galactic rotation curves is then obtained \cite{Mannheim2010b,Mannheim2010c,O'Brien2011}. In the fits the only free parameters are the galactic mass to light ratios (one free parameter per galaxy) that normalize the strength of  $B_{\rm LOC}(r)$ to the measured galactic luminosities, with everything else being universal. On defining $\gamma^*$ as the coefficient in (\ref{P130}) for one solar mass of material, and with $\beta^*=1.48\times 10^{5}~{\rm cm}$ of the Sun already being known, from the fitting the three needed parameters of the conformal theory are found to be given by 
\begin{equation}
\gamma^*=5.42\times 10^{-41}~{\rm cm}^{-1},\qquad \gamma_0=3.06\times
10^{-30}~{\rm cm}^{-1},\qquad \kappa =9.54\times 10^{-54}~{\rm cm}^{-2}.
\label{P137}
\end{equation} 
These numbers confirm that $\gamma_0$ is indeed of cosmological magnitude and that $\kappa$ is indeed associated with a somewhat smaller, inhomogeneous cluster-sized scale, with the success of the fitting suggesting that the use of dark matter in galaxies is nothing more than an attempt to describe global physics effects in purely local galactic terms. For our purposes in this paper, we note that the use of (\ref{P136}) not only involves  a detailed interplay between the homogeneous cosmological background and the inhomogeneities in it, through explicit determination of the parameters $\gamma_0$ and $\kappa$, galactic rotation curves are directly measuring both the spatial curvature of the background cosmology and the moment integral given in (\ref{P132}) that is associated with the inhomogeneities in it. Once the study of this paper is extended to matter fluctuations, one will be able to ascertain the degree to which such fluctuations are aware of either of these $\gamma_0$ and $\kappa$ scales.

\subsection{Implications for Inhomogeneities}

As well as being related to the growth of inhomogeneities, the $\kappa$ term also has implications for inhomogeneities even after they have been formed. The inhomogeneous structure of systems such as clusters of galaxies for instance can be explored observationally by two kinds of probes, one an interior one and the other an exterior one. The interior probe involves measuring galaxy kinematics and X-ray kinematics, while the exterior probe involves measuring lensing by clusters. For the interior case we need to use (\ref{P129}) for points within the cluster, and for lensing we need to use  (\ref{P129}) for points exterior to the cluster, and in both cases we need to include the global effect due to the rest of the material in the Universe. Since previous applications of conformal gravity to clusters (velocity dispersions \cite{Mannheim1995}, X-rays in clusters \cite{Horne2006,Diaferio2009}, and lensing  \cite{Pireaux2004,Sultana2010}) did not include such a global effect,  studies of its possible impact on clusters and also on gamma ray bursters (see \cite{Schaefer2003,Speirits2007} and the encouraging recent analysis of \cite{Diaferio2011}) could be instructive. 

The gravitational lensing issue is of particular interest since it was noted in \cite{Walker1994} and \cite{Edery1998} that if one uses the standard gravitational bending formula due to a point source but with the metric given in (\ref{P130}), for a positive value of the $\gamma$ parameter (as per (\ref{P137})) one finds bending away from the source (defocusing) rather than toward it. (For an explanation of how a linear potential could be attractive for non-relativistic motions and repulsive for relativistic ones see \cite{Mannheim2000a}.) However, in a recent study of gravitational bending in a de Sitter geometry, Rindler and Ishak \cite{Rindler2009,Ishak2010} pointed out that because the de Sitter geometry is not asymptotically flat one could not use the standard textbook gravitational bending formula, and they instead provided an alternate prescription. The same concerns equally apply to the linear potential case as well, as it too is not asymptotically flat, with the study of \cite{Sultana2010} finding that the revised amount of bending away from the source was then completely negligible. Still left to be done is to see whether the inclusion of material external to the source as embodied in the $\gamma_0$ and $\kappa$ terms could then produce bending toward the source, bending that is commonly attributed to dark matter.

\subsection{Implications for Second-order Fluctuations}

While we have shown that we can use the underlying conformal symmetry of the conformal gravity theory to monitor the behavior of first-order gravitational fluctuations around backgrounds that are conformal to flat even if the fluctuations themselves are not, we can extend the analysis to second-order fluctuations as well. In fact, just as with the first-order fluctuation equations, the second-order fluctuation equations are also found \cite{Mannheim2012} to only depend on the traceless part of the fluctuation. For instance, for transverse-traceless fluctuations around a flat background the term in the $I_{\rm W}$ action of (\ref{P1}) that is second order in the fluctuation $h_{\mu\nu}$  is found to be of the form  \cite{Mannheim2009}
\begin{equation}
I_{\rm W}=-\frac{\alpha_g}{2}\int d^4x \partial_{\alpha}\partial^{\alpha} K_{\mu\nu}\partial_{\beta}\partial^{\beta} K^{\mu\nu},
\label{P138}
\end{equation}
to indeed only depend on the traceless $K_{\mu\nu}$.

Given (\ref{P138}), by functional variation with respect to the metric  we can obtain the contribution of a transverse-traceless $K_{\mu\nu}$ to the second-order $W_{00}(2)$ and thereby determine the energy carried by a gravity wave. To be specific, we note that  for first-order tensor fluctuations around a flat background  the general solution based on (\ref{P74}) takes the form
\begin{eqnarray}
K_{\mu\nu}(x)&=&\frac{1}{2(-\alpha_g)^{1/2}}\sum_{i=1}^2\int \frac{d^3k}{(2\pi)^{3/2}(\omega_k)^{3/2}}\bigg{[}
A^{(i)}(\bar{k})\epsilon^{(i)}_{\mu\nu}(\bar{k})e^{ik\cdot x}
+i\omega_kB^{(i)}(\bar{k})\epsilon^{(i)}_{\mu\nu}(\bar{k})(n\cdot x)e^{ik\cdot x}\
\nonumber\\
&+&\hat{A}^{(i)}(\bar{k})\epsilon^{(i)}_{\mu\nu}(\bar{k})e^{-ik\cdot x}-
i\omega_k\hat{B}^{(i)}(\bar{k})\epsilon^{(i)}_{\mu\nu}(\bar{k})(n\cdot x)e^{-ik\cdot x}\bigg{]}
\label{P139}
\end{eqnarray}
when written in a massless plane wave basis with $\omega_k=|\bar{k}|$. In this solution $K_{\mu\nu}(x)$ is expressed in terms of operators $A^{(i)}(\bar{k})$, $\hat{A}^{(i)}(\bar{k})$, $B^{(i)}(\bar{k})$ and  $\hat{B}^{(i)}(\bar{k})$, two transverse-traceless polarization tensors $\epsilon^{(i)}_{\mu\nu}(\bar{k})$ ($i=1,2)$, both of which are normalized to $\epsilon_{\alpha\beta}\epsilon^{\alpha\beta}=1$, and a unit timelike reference vector $n^{\mu}=(1,0,0,0)$. If we evaluate the contribution of this $K_{\mu\nu}$ to the second-order perturbation correction to $W_{00}$ in a background that is flat, we obtain \cite{Mannheim2009}
\begin{equation}
-4\alpha_g\int d^3 x W_{00}(2)
=\sum_i\int d^3k\omega_k\left[\hat{A}^{(i)}(\bar{k})B^{(i)}(\bar{k})+A^{(i)}(\bar{k})\hat{B}^{(i)}(\bar{k}))+2\hat{B}^{(i)}(\bar{k})B^{(i)}(\bar{k})\right].
\label{P140}
\end{equation}

Inspection of (\ref{P140}) shows that if we drop the asymptotically-non-flat $B^{(i)}(\bar{k})$ modes, then since there are no $\hat{A}^{(i)}(\bar{k})A^{(i)}(\bar{k})$ cross terms in (\ref{P140}), we will be left with $\int d^3 x W_{00}(2)=0$, with the $A^{(i)}(\bar{k})$ modes themselves thus carrying no energy. Since, as shown in Sec. (4.3), the $A^{(i)}(\bar{k})$  type modes are also solutions to a second-order wave equation, we see \cite{Deser2007,Bouchami2008,Mannheim2009,Fabri2011} that in the fouth-order conformal gravity theory those  fourth-order gravity wave modes that are also solutions to standard second-order Einstein gravity do not carry any energy in the fourth-order case even though they of course do carry energy in the second-order case. Thus, while the $A_{\mu\nu}$ type gravitational wave is a solution to both the second- and fourth-order theories, we see   that in the conformal case it is not able to carry any energy at all. As noted in \cite{Mannheim2009}, it is only the $B_{\mu\nu}$ type solution in (\ref{P74}) and (\ref{P140}) that carries energy. 

The form for  $\int d^3 x W_{00}(2)$ given in (\ref{P140}) is to be anticipated given the conformal gravity zero-energy theorem of Boulware, Horowitz and Strominger 
\cite{Boulware1983}, who showed that if one restricts to asymptotically flat solutions (i.e. effectively keep $A_{\mu\nu}$ but not $B_{\mu\nu}$), the total energy carried by a gravity wave would be zero. However, the theorem imposes no constraint on non-asymptotically-flat modes, and if we drop the $A^{(i)}(\bar{k})$ modes from $\int d^3 x W_{00}(2)$, the contribution of the $B^{(i)}(\bar{k})$ modes will not then be required to vanish.

Given the above result for fluctuations around a flat background, we can now use the techniques developed in the first-order case to infer that if  we fluctuate around  a background that is conformal to flat, because of (\ref{P8}) we will find that the contribution of the conformally transformed $\bar{A}_{\mu\nu}$ type modes to the conformally transformed $\int d^3 x \bar{W}_{00}(2)$ will be zero too. Thus even though the fluctuations around a background that is conformal to flat are not themselves conformal to flat, such fluctuations will always contain some modes that carry no energy in second perturbative order. To identify the particular modes involved for fluctuations around a de Sitter background for instance, we note that on comparing (\ref{P82}) with (\ref{P139}), we see that solutions to  $[\bar{\nabla}_{\alpha}\bar{\nabla}^{\alpha}-2H^2]\bar{h}_{\mu\nu}=0$ have $B^{(i)}(\bar{k})=-A^{(i)}(\bar{k})$, $\hat{B}^{(i)}(\bar{k})=-\hat{A}^{(i)}(\bar{k})$, and thus cause $\int d^3 x \bar{W}_{00}(2)$ to vanish. de Sitter background $[\bar{\nabla}_{\alpha}\bar{\nabla}^{\alpha}-2H^2]\bar{h}_{\mu\nu}=0$ type modes thus  carry energy in standard gravity but not in the conformal case. While beyond the scope of the present paper, differences such as these between second-order and fourth-order gravity waves could eventually enable one to distinguish between the two theories.

\subsection{Relating Einstein Gravity and Conformal Gravity}

While this paper has concentrated on the conformal gravity theory and compared its predictions and structure with that found in Einstein gravity, there are some recent indications 
\cite{'tHooft2010a,'tHooft2010b,'tHooft2011,Maldacena2011} that the two theories may be related in some way. The idea that there might be a possible connection between the two gravitational theories is an old one, with authors such as Adler \cite{Adler1982}, Zee \cite{Zee1983}, and Hoyle and Narlikar \cite{Hoyle1964} having explored either the $\int d^4x\, (-g)^{1/2}C_{\lambda\mu\nu\kappa}C^{\lambda\mu\nu\kappa}$ action or a conformally coupled scalar field theory. 

Renewed interest in the possible existence of a connection between the two gravitational theories was recently generated by 't Hooft in a series of papers 
\cite{'tHooft2010a,'tHooft2010b,'tHooft2011} in which he reformulated the Einstein gravity path integral by integrating not over the individual $g_{\mu\nu}(x)$ metric components themselves but by first integrating over the determinant of the metric, this essentially being equivalent to treating the metric conformal factor as an independent dynamical degree of freedom. On doing the path integral this way 't Hooft obtained a dimensionally regularized effective action of the form 
\begin{equation}
I_{\rm EFF}=\frac{C}{120}\int d^4x (-g)^{1/2}[R^{\mu\nu}R_{\mu\nu}-{1 \over 3}(R^{\alpha}_{\phantom{\alpha}\alpha})^2],
\label{P141}
\end{equation}
where $C=1/8\pi^2(4-D)$ in spacetime dimension $D$. In (\ref{P141}) we immediately recognize 
the conformal gravity action given in (\ref{P1}), with the conformal gravity action thus naturally being generated in a theory in which one starts with the Einstein-Hilbert action alone.

However as also noted by 't Hooft and reemphasized in \cite{Mannheim2011}, if one instead starts not with a gravity theory at all but with a fermionic theory in which the fermion kinetic energy operator is coupled to arbitrary external gravitational and electromagnetic fields according to
\begin{equation}
I_{\rm M}=-\int d^4x (-g)^{1/2}\bar{\psi}(x)\gamma^{\mu}(x)[i\partial_{\mu}+i\Gamma_{\mu}(x)+A_{\mu}(x)]\psi(x),
\label{P142}
\end{equation}
then without making any reference at all to the form of the action in the gravitational sector, a path integration over the fermionic fields yields a dimensionally regularized effective action of the form 
\begin{equation}
I_{\rm EFF}=\int d^4x (-g)^{1/2}C\left[\frac{1}{20}[R^{\mu\nu}R_{\mu\nu}-{1 \over 3}(R^{\alpha}_{\phantom{\alpha}\alpha})^2]
+\frac{1}{3}F_{\mu\nu}F^{\mu\nu}\right],
\label{P143}
\end{equation}
where $C$ is precisely the same constant as the one that appears in (\ref{P141}). From (\ref{P143}) we see that not only do we generate the conformal gravity action, we see that it is the conformal gravity action that should be thought of as being on an equal footing with the Maxwell action rather than the Einstein-Hilbert one. Since a fermionic path integral over the completely standard and mundane action in (\ref{P142}) must appear in any theory of gravity (a path integration that is equivalent to a single closed fermion loop Feynman diagram), one is never free to study gravity theories in which the conformal gravity action is not included. One can argue only over the possible strength of such terms, with the coefficient $C$ actually being infinite in $D=4$ (just as it is for the Maxwell term \cite{footnote3}). Since such conformal gravity terms have to be there, one needs to explore their possible implications and determine their strength from experiment, with the material presented in this paper being an initial step toward that goal.

While the objective of 't Hooft was to generate conformal gravity starting from Einstein gravity, the objective of Maldacena \cite{Maldacena2011} was to do the converse, namely to generate Einstein gravity starting from the conformal theory. Now this had also been the objective of Adler, of Zee, and of Hoyle and Narlikar, who looked to descend from conformal gravity to Einstein gravity via spontaneous breakdown of the conformal symmetry. However, the approach of Maldacena  is somewhat different. Specifically, he starts by noting that in the pure gravity sector, any geometry in which $G_{\mu\nu}+\Lambda g_{\mu \nu}$ is zero ($\Lambda$ being a zero or non-zero constant) will be a geometry in which $W_{\mu\nu}$ is zero too. Hence solutions to Einstein gravity that obey $G_{\mu\nu}+\Lambda g_{\mu \nu}=0$ will equally be solutions to conformal gravity too. However, as indicated for instance in (\ref{P74}) and (\ref{P130}), the conformal theory has other solutions too. Thus if one could find a way to eliminate these extra solutions, in the pure gravity sector one would be left with the solutions to Einstein gravity alone. To this end Maldacena studied de Sitter space and imposed an asymptotic  future boundary condition that is to exclude solutions that grow linearly (or analogously) in time, to thereby remove the $B_{\mu\nu}$ type solutions given in (\ref{P74}) but not the $A_{\mu\nu}$ type solutions. In this way one is then left only with the plane wave solutions that also appear in the standard Einstein theory \cite{footnote4}. 

To conclude, we note that one can delineate three possible options for gravitational theories: pure Einstein gravity, pure conformal gravity, or an Einstein gravity that is related to conformal gravity. Further analysis of the type developed in this paper could enable one to distinguish between the various cases.


\begin{thebibliography}{99}




\bibitem{Mannheim2006} P.~D.~Mannheim,~Prog.~Part.~Nucl.~Phys.~{\bf 56},~340~(2006). 

\bibitem{footnote0} By restricting to polynomial Lagrangians we exclude locally conformal invariant actions such as $\int d^4x (-g)^{1/2}[C_{\lambda\mu\nu\kappa}C^{\lambda\mu\sigma\tau}C^{\nu\kappa}_{\phantom{\lambda\mu}\sigma\tau}]^{2/3}$ that are based on fractional powers.

\bibitem{Mannheim1990} P.~D.~Mannheim,~Gen.~Rel.~Gravit.~{\bf 22}, 289 (1990).

\bibitem{Mannheim2008} P.~D.~Mannheim,~{\it Dynamical symmetry breaking and the cosmological constant problem}, Proceedings of the 34th International Conference in High Energy Physics (ICHEP08), Philadelphia, 2008, eConf C080730.~(arXiv:0809.1200 [hep-th]) 

\bibitem{Mannheim2009} P.~D.~Mannheim,~Gen.~Rev.~Gravit.~{\bf 43}, 703 (2011).

\bibitem{Mannheim2010a} P.~D.~Mannheim,~Mod.~Phys.~Lett.~A {\bf 26}, 2375 (2011).

\bibitem{Bender2008} C.~M.~Bender and P.~D.~Mannheim,~Phys.~Rev.~Lett.~{\bf 100},~110402 (2008);~Phys.~Rev.~D {\bf 78}, 025022 (2008);~J.~Phys.~A {\bf 41}, 304018 (2008).

\bibitem{Mannheim2011} P.~D.~Mannheim,~Found.~Phys.~{\bf 42}, 388 (2012).

\bibitem{footnote00} The discussion of the conformal trace anomaly in the conformal theory is somewhat different from its discussion in standard gravity. Specifically, because Einstein gravity is not renormalizable, in standard gravity one uses a semi-classical approach in which a classical Einstein tensor is coupled to matrix elements of a quantum-mechanical matter field energy-momentum tensor $T^{\mu\nu}$. This matter field $T^{\mu\nu}$ possesses a radiatively-induced trace anomaly, and efforts to produce a semi-classical theory in which a classical Einstein tensor would couple to an anomaly-free matter field $T^{\mu\nu}$ have yet to succeed. In conformal gravity the matter field sector discussion is not changed and the matter field $T^{\mu\nu}$ still has a radiatively-induced trace anomaly. However, what changes is that the gravity sector now is renormalizable, and thus one now can discuss a theory of gravity in which both the matter and gravity sectors can be treated quantum-mechanically. When one does this, the gravity sector then also acquires a radiatively-induced trace anomaly, but, as  shown in \cite{Mannheim2011}, the gravitational equations of motion given in (\ref{P3}) require that the anomalies in the gravity and matter sectors cancel identically. Thus even while both the gravitational $W^{\mu\nu}$ and the matter field $T^{\mu\nu}$ that appear in (\ref{P3}) separately obey conformal Ward identities that are violated by radiative corrections, the stationarity condition (\ref{P3}) that relates them is not itself a conformal Ward identity, and thus it itself acquires no radiatively-induced anomalous terms. In consequence, the $W^{\mu\nu}$ and matter field $T^{\mu\nu}$ trace anomaly terms must mutually cancel each other identically, with the combination $4\alpha_g W^{\mu\nu}-T^{\mu\nu}$ that appears in (\ref{P3}) thus being anomaly free. That this cancelation occurs is due to the fact that (\ref{P3}) automatically fixes the wave function renormalization constant of the gravitational field to be that function of the wave function renormalization constants of the matter fields that causes infinities in the gravitational and matter sectors of (\ref{P3}) to mutually cancel each other. With the infinite parts of (\ref{P3}) thus canceling each other out, one can therefore use the residual finite part of (\ref{P3}) to study observational phenomena such as the ones explored in this paper.

\bibitem{Mannheim1997} P.~D.~Mannheim,~Ap.~J.~{\bf 479}, 659 (1997).

\bibitem{Mannheim2010b} P.~D.~Mannheim and J.~G.~O'Brien,~Phys.~Rev.~Lett.~{\bf 106}, 121101 (2011).

\bibitem{Mannheim2010c} P.~D.~Mannheim and J.~G.~O'Brien,~{\it Fitting galactic rotation curves with conformal gravity and a global quadratic potential},~Phys.~Rev.~D, in press. (arXiv:1011.3495 [astro-ph.CO])

\bibitem{O'Brien2011} J.~G.~O'Brien and P.~D.~Mannheim,~Mon.~Not.~Roy.~Ast.~Soc. {\bf 421}, 1273 (2012).

\bibitem{Mannheim2003} P.~D.~Mannheim, Intl.~Jour.~Mod.~Phys.~D {\bf 12}, 893 (2003).

\bibitem{Mannheim1989} P.~D.~Mannheim and D.~Kazanas,~Ap.~J.~{\bf 342}, 635 (1989).

\bibitem{Mannheim2005} P. D. Mannheim, {\it Brane-Localized Gravity}
(World Scientific, New Jersey, 2005).

\bibitem{footnote000} While foreign to second-order derivative theories, fluctuation modes that are power-behaved in time are characteristic of higher-derivative theories. Such modes have the unusual feature of not being eigenstates of  $\partial/\partial t$, and thus in the quantized version of the theory they are associated with wave functions that are not energy eigenstates. In consequence, the quantized Hamiltonian of a fourth-order  theory such as conformal gravity cannot have a complete set of eigenstates and must thus be neither diagonalizable nor Hermitian. It was through recognizing that there was such a lack of Hermiticity that Bender and Mannheim \cite{Bender2008} were able to construct a unitary quantum-mechanical realization of the fourth-order theory that was free of negative norm states. As we note below in our discussion of fluctuations in RW cosmologies, having fluctuation growth that is linear in time enables conformal cosmology to produce RW fluctuations that grow far more rapidly than the ones that occur in the standard second-order theory. In our study below of the geometry associated with a static source we shall encounter potentials that grow linearly with distance. As such they are the spatial counterparts of a linear in time growth. 

\bibitem{Mannheim1991a} P.~D.~Mannheim,~Ap.~J.~{\bf 391}, 429 (1992).


\bibitem{Mannheim1998} P.~D.~Mannheim,~Phys.~Rev.~D {\bf 58}, 103511 (1998).

\bibitem{Mannheim2001} P.~D.~Mannheim,~Ap.~J.~{\bf 561}, 1 (2001).


\bibitem{Mannheim2000a} P.~D.~Mannheim,~Found.~Phys.~{\bf 30}, 709 (2000). 

\bibitem{Guth1981}  A.~H.~Guth,~Phys.~Rev.~D {\bf 23}, 347 (1981).

\bibitem{Copi2010} C.~J.~Copi,~D.~Huterer,~D.~J.~Schwarz and G.~Starkman,~Adv.~Astron.~{\bf 2010},~847541~(2010).


\bibitem{footnote1} In the same vein we note that for an $a(t)$ that behaves as $t^n$, $\dot{a} \sim nt^{n-1}$ and $\ddot{a}(t)\sim n(n-1)t^{n-2}$. Thus for cosmologies with $n \geq 1$  there is no horizon problem, no initial singularity, and no epoch in which the Universe decelerates. However, for cosmologies with $0<n<1$ there is a horizon problem, there is an initial singularity, and there is no epoch in which the Universe accelerates. Since it is the presence of an initial singularity in the standard Friedmann  evolution equations that leads to the flatness problem [$\dot{a}^2+K=\dot{a}^2\Omega(t)$ requires that $\Omega(t_i)=1$ if $\dot{a}(t_i)=\infty$ no matter what the value of $K$], we see that the standard model horizon, flatness and acceleration problems all have a common origin, namely that in the standard model $n$ is less than one.

\bibitem{footnote2} As noted in \cite{Mannheim2000a}, for the typical $a(t,\alpha=0,K<0)$ for instance, the horizon size integral is given by $(-K)^{1/2}d_{\rm H}(t)={\rm arc sinh}[(T^2_{\rm MAX}/T^2(t)-1)^{1/2}]$, to thus be huge at recombination. For a cosmology with this $a(t,\alpha=0,K<0)$, or equally for a $K<0$ cosmology with $\alpha>0$, there is no initial singularity, with the Universe expanding from a large but finite initial temperature. [As can be seen from (\ref{P121}) and (\ref{P122}), the conformal cosmology acts just like a standard Friedmann cosmology except that it has an effective Newton constant given by the negative definite (i.e. cosmologically repulsive) $G_{\rm EFF}=-3/4\pi S_0^2$. As noted in \cite{Mannheim1991a}, cosmologies with $G_{\rm EFF}<0$ have no initial singularity.] In addition we note that in the $\alpha=0$, $K<0$  conformal cosmology the deceleration parameter is given by \cite{Mannheim2001} $q(t,\alpha=0,K<0)=-2A/K^2S_0^2t^2 = -T^2(t)/(T_{\rm MAX}^2-T^2(t))$, to thus naturally be negative definite and lie within the range $-1 \leq q(t)\leq 0$ for all $T^2(t)\leq T^2_{\rm MAX}/2$. Thus neither $q(t,\alpha=0,K<0)$, nor equally $q(t,\alpha>0,K<0)$, which also has to lie within the range $-1 \leq q(t)\leq 0$ at late times no matter how large $\alpha$ might be \cite{Mannheim2001}, ever require any fine tuning. As noted in \cite{footnote1}, the horizon, flatness and acceleration problems all stand or fall together, with all of them being simultaneously solved in the conformal case if $K$ is negative. Despite this, because of its evolution rate,  the conformal theory does have a shortcoming in that it does not produce enough primordial deuterium  \cite{Knox1993,Elizondo1994,Lohiya1999,Sethi1999}. As is conventional, these studies were all made in a strictly homogeneous cosmology.  However, as noted in \cite{Lohiya1999} and \cite{Mannheim2001}, it is also possible to produce deuterium inhomogeneously,  just as inhomogeneities first begin to form in the Universe. In this paper we have taken the first step needed to address this issue, an issue that is central to the viability of the conformal theory.

\bibitem{Knox1993} L.~Knox and  A.~Kosowsky, {\it  Primordial nucleosynthesis in conformal Weyl gravity}, November 1993. (astro-ph/9311006)  

\bibitem{Elizondo1994} D.~Elizondo and G.~Yepes,~Ap.~J.~
{\bf 428}, 17 (1994).

\bibitem{Lohiya1999} D.~Lohiya,~A.~Batra,~S.~Mahajan and A.~
Mukherjee,  {\it Nucleosynthesis in a simmering universe}, February 1999. (nucl-th/9902022)

\bibitem{Sethi1999} M.~Sethi,~A.~Batra and D.~Lohiya, Phys.~Rev.~D
{\bf 60}, 108301 (1999).


\bibitem{Mannheim1994} P.~D.~Mannheim and D.~Kazanas,~Gen.~Rel.~Gravit.~{\bf 26}, 337 (1994).

\bibitem{Mannheim1995} P.~D.~Mannheim,~{\it Linear potentials in galaxies and clusters of galaxies},~April~1995. (astro-ph/9504022)

\bibitem{Horne2006} K.~Horne, Mon.~Not.~R.~Astron.~Soc.~{\bf 369}, 1667 (2006).



\bibitem{Diaferio2009} A.~Diaferio and L.~Ostorero, Mon.~Not.~R.~Astron.~Soc.~{\bf 393}, 215 (2009).

\bibitem{Pireaux2004} S.~Pireaux, Class.~Quant.~Gravit. {\bf 21}, 1897, 4317 (2004).


\bibitem{Sultana2010} J.~Sultana and D.~Kazanas, Phys.~Rev.~D~{\bf 81}, 127502 (2010).

\bibitem{Schaefer2003} B.~E.~Schaefer, Ap.~J.~{\bf 583}, L67 (2003); Ap.~J.~ {\bf 660}, 16 (2007).

\bibitem{Speirits2007} F.~C.~Speirits,~M.~A.~Hendry and A.~Gonzalez,~Phil.~Trans.~Roy.~ Soc.~A: {\bf  365}, 1395 (2007).

\bibitem{Diaferio2011} A.~Diaferio,~L.~Ostorero,~V.~F.~Cardone,~Jour.~Cos.~Astrop.~Phys.~{\bf 10}, 008 (2011).

\bibitem{Walker1994} M.~A.~Walker, Ap.~J. {\bf 430}, 463 (1994).

\bibitem{Edery1998} A.~Edery and M.~B.~Paranjape,~Phys.~Rev.~D {\bf 58}, 024011 (1998).

\bibitem{Rindler2009} W.~Rindler and M.~Ishak,~Phys.~Rev.~D {\bf 76}, 043006 (2007).

\bibitem{Ishak2010} M.~Ishak and W.~Rindler,~Gen.~Rel.~Gravit.~{\bf 42}, 2247 (2010).

\bibitem{Mannheim2012} P.~D.~Mannheim and M.~B.~Paranjape (unpublished).

\bibitem{Deser2007}  S.~Deser and B.~Tekin, Phys.~Rev.~D {\bf 75}, 084032 (2007).

\bibitem{Bouchami2008} J.~Bouchami and M.~B.~Paranjape, Phys.~Rev.~D {\bf 78}, 044022 (2008). 
 
\bibitem{Fabri2011} L.~Fabbri and M.~B.~ Paranjape,~Phys.~Rev.~D {\bf 83}, 104046 (2011).

\bibitem{Boulware1983}  D.~G.~Boulware,~G.~T.~Horowitz, and A.~Strominger, Phys.~Rev.~Lett.~{\bf 50}, 1726 (1983).



\bibitem{'tHooft2010a} G.~'t~Hooft,  {\it Probing the small distance structure of canonical quantum gravity using the conformal group}, September 2010. (arXiv:1009.0669 [gr-qc])

\bibitem{'tHooft2010b} G.~'t~Hooft,  {\it The conformal constraint in canonical quantum gravity}, November 2010. (arXiv:1011.0061 [gr-qc])


\bibitem{'tHooft2011} G.~'t~Hooft, ~Found.~Phys.~{\bf 41}, 1829 (2011).

\bibitem{Maldacena2011} J.~Maldacena, {\it Einstein gravity from conformal gravity}, May 2011. (arXiv:1105.5632 [hep-th])
 
 
 \bibitem{Adler1982} S.~L.~Adler,~Rev.~Mod.~Phys.~{\bf 54}, 729 (1982).
 
 \bibitem{Zee1983} A.~Zee,~Ann.~Phys.~(N.~Y.)~{\bf 151}, 431 (1983).

 \bibitem{Hoyle1964} F.~Hoyle and J.~V.~Narlikar,~Proc.~Roy.~Soc.~A.~{\bf 282}, 191 (1964).
 
 
 \bibitem{footnote3} For the Maxwell term one can remove the infinity in (\ref{P143}) by having it renormalize an intrinsic Maxwell term that is present in the fundamental starting action. If one  thus includes an intrinsic $-\alpha_g\int d^4x\, (-g)^{1/2}C_{\lambda\mu\nu\kappa}
C^{\lambda\mu\nu\kappa}$ term in the starting action too, then the gravitational sector infinity in (\ref{P143}) can also be removed by renormalization.


 \bibitem{footnote4} While the $A_{\mu\nu}$ type modes are solutions to both the standard gravity $G_{\mu\nu}+\Lambda g_{\mu \nu}=0$ and the conformal gravity $W_{\mu\nu}=0$ in first-perturbative order, as we noted above, unlike in standard gravity, in conformal gravity the $A_{\mu\nu}$ type modes do not carry energy in second-perturbative order.  




\end{thebibliography}
\end{document}